\documentclass[11pt,preprint]{aastex}

\begin{document}
\title{Measuring the Slope of the Dust Extinction Law and the Power Spectrum
of Dust Clouds Using Differentially-Reddened Globular Clusters}

\author{J.\ Melbourne and P.\ Guhathakurta}
\affil{UCO/Lick Observatory, Department of Astronomy and Astrophysics, 
University of California at Santa Cruz, 1156 High Street, Santa Cruz,
California~95064}
\email{jmel, raja@ucolick.org}

\begin{abstract}
We present three~methods for measuring the slope of the Galactic dust
extinction law, $R_V$, and a method for measuring the fine-scale
structure of dust clouds in the direction of differentially-reddened globular
clusters.  We apply these techniques to $BVI$ photometry of stars in the
low-latitude Galactic globular cluster NGC~4833 which displays
spatially-variable extinction/reddening about a mean
$\langle{A_V}\rangle\approx1$.  An extensive suite of Monte Carlo simulations
is used to characterize the efficacy of the methods.  The essence of the
first two~methods is to determine, for an assumed value of $R_V$, the {\it
relative\/} visual extinction $\delta{A_V}$ of each cluster horizontal branch
(HB) star with respect to an empirical HB locus; the locus is derived from
the color-magnitude diagram (CMD) of a subset of stars in a small region near
the cluster center for which differential extinction/reddening are relatively
small.  A star-by-star comparison of $\delta{A_V}$ from the ($B-V$,~$V$) CMD
with that from the ($V-I$,~$V$) CMD is used to find the optimal $R_V$.  In
the third method, $R_V$ is determined by minimizing the scatter in the HB in
the ($B-V$,~$V$) CMD after correcting the photometry for extinction and
reddening using the Schlegel, Finkbeiner, \& Davis (1998) dust maps.  The
weighted average of the results from the three~methods gives $R_V=3.0\pm0.4$
for the dust along the line of sight to NGC~4833.  The fine-scale structure
of the dust is quantified via the difference, $(\Delta{A_V})_{ij}\equiv
(\delta{A_V})_i-(\delta{A_V})_j$, between pairs of cluster HB stars
($i$,\,$j$) as a function of their angular separation $r_{ij}$.  The variance
(mean square scatter) of $(\Delta{A_V})_{ij}$ is found to have a power-law
dependence on angular scale: ${\rm var}(r)\propto{r}^\beta$, with
$\beta=+0.9\pm0.1$.  This translates into an angular power spectrum
$P(\kappa)\propto\kappa^\alpha$, with the index $\alpha=-1.9\pm0.1$ for
$r\sim1'$--$5'$, where $\kappa\equiv1/r$.  The dust angular power spectrum on
small scales (from optical data) matches smoothly onto the larger-scale power
spectrum derived from Schlegel et~al.'s far-infrared map of the dust thermal
emission.
\end{abstract}

\keywords{Galaxy: globular clusters: individual NGC~4833 --- ISM: dust,
extinction --- ISM: structure}

\section{Introduction}

Scattering, absorption, and reradiation of photons by dust grains affect the
propagation of starlight and play a key role in regulating the energy balance
in the interstellar medium.  Studies of these processes have provided a great
deal of insight into the physical and chemical properties of dust grains (for
reviews of dust properties, see Mathis 1990a; Witt 2003).  It is also
important to disentangle the effects of intervening dust from many kinds of
astronomical observations, ranging from photometry and spectroscopy of
Galactic stars and external galaxies to mapping small-scale anisotropy in the
Cosmic Microwave Background Radiation (CMBR).

\subsection{Reddening Law and Dust Grain Properties}

The degree of absorption and scattering by dust depends, in general, on the
wavelength of the incident radiation (Whitford 1958).  This dependence can be
quantified in terms of the normalized dust extinction law $A_\lambda/A_V$ (or
reddening law), where extinction is the sum of absorption plus scattering.
The ratio of total to selective extinction, $R_V\equiv{A_V}/E(B-V)$, is a
commonly-used measure of the slope of the extinction/reddening law.  Using
data along many different Galactic sight-lines, Cardelli, Clayton, \& Mathis
(1989, hereafter CCM) found a tight correlation between the overall shape of
the reddening law and the slope $R_V$.  CCM devised an $R_V$-based,
one-parameter family of empirical fitting functions to characterize the
observed range of extinction law shapes; this parameterization was later
refined by Fitzpatrick (1999).

Galactic dust displays a wide range of behavior.  While diffuse interstellar
dust is observed to have $R_V=3.1$ {\it on average\/} (Savage \& Mathis 1979;
CCM), this value is by no means universal.  Significant deviations from this
canonical $R_V=3.1$ value (and corresponding differences in the overall shape
of the extinction/reddening law) are known to exist for a variety of
interstellar dust clouds, and these variations appear to be generally
correlated with environment (Fitzpatrick 1999).  For example, $R_V$ has been
found to range from 4 to 5 in dense molecular clouds (Mathis 1990b; Larson,
Whittet, \& Hough 1996; Whittet et~al.\ 2001), whereas there are indications
that more diffuse, high-latitude cirrus clouds may have $R_V$ values as small
as $\sim2$ (Fitzpatrick \& Massa 1990; Larson et~al.\ 1996; Szomoru \&
Guhathakurta 1999).  Studies based on stellar photometry from the Optical
Gravitational Lens Experiment (OGLE) and MAssive Compact Halo Objects (MACHO)
projects point to $R_V$ being substantially smaller than~3 in the direction
of the Galactic bulge (Popowski 2000; Popowski, Cook, \& Becker 2003; Udalski
2003; Sumi 2004; Popowski 2004).  Barbaro et~al.\ (2001) find departures from
the CCM parameterization in the ultraviolet portion of the Galactic
extinction law.  It is worth noting that the above studies sample dust at a
wide range of distances from the Sun: some probe the diffuse interstellar
medium in its immediate vicinity whereas others, especially those at low
Galactic latitudes, can probe dust at much larger distances well outside the
Solar neighborhood (e.g.,~the RCrA cloud studied by Szomoru \& Guhathakurta
1999).

Some studies have suggested that $R_V$ variations are tied to variations in
the size distribution of dust grains from one line of sight to another.  In
diffuse cirrus clouds, a relative abundance of small grains might explain the
steep rise of the extinction curve into the ultraviolet (CCM; Larson et~al.\
1996).  By contrast, a high abundance of large grains, which grow readily by
coagulation in dense molecular clouds, may explain the larger than average
values of $R_V$ observed in these regions (Whittet et~al.\ 2001).  Whittet
et~al.\ suggest that, in dense molecular clouds, $R_V$ remains close to the
standard value of~3.1 except for lines of sight with unusually high
extinction ($A_V\ge3$).  This may indicate a more complicated relationship
between $R_V$ and $A_V$.  Moreover, it has been argued that chemical
composition can also play a role in determining the shape of the extinction
law (Rhoads, Malhotra, \& Kochanski 2004).

To improve our understanding of the dependence of $R_V$ on environment,
measurements are necessary along many sight-lines through a broad range of
cloud types.  Clusters with differential extinction/reddening have long been
used to measure the $R_V$ of intervening dust (Mihalas \& Routly 1968).  The
traditional method of spectral typing individual cluster stars (e.g.,~on the
Morgan-Keenan system), in conjunction with photometric measurements, allows
for a direct and precise measurement of extinction and reddening.  However,
this method of measuring $R_V$ requires high-quality, flux-calibrated spectra
and a good understanding of the effects of metallicity on the energy output
of stars.  Our study focuses instead on the use of broad-band photometry in
light of the fact that an extensive suite of photometric data sets is
currently available.

In this paper, we propose three~methods for measuring the $R_V$ of dust in
the foreground of differentially-reddened globular clusters.  The first
two~methods are closely related to each other and rely on accurate three-band
optical photometry of cluster stars.  The third relies on two-band optical
photometry and a map of the dust thermal emission (Schlegel, Finkbeiner, \&
Davis 1998, hereafter SFD) and is particularly well suited to wide-field
data.  All three~methods are based on the premise that the width of the blue
horizontal branch (HB) is minimized when the appropriate $R_V$ value is used
to correct the photometry for extinction and reddening.  The methods are
applied to NGC~4833, a low-latitude Galactic globular cluster with variable
extinction across its face.  Realistic Monte Carlo simulations of the cluster
data set are used to estimate the accuracy with which $R_V$ can be recovered.

\subsection{``Cirrus''---The Rich Texture of Interstellar Dust}

Nearly two decades ago, Low et~al.\ (1984) noticed diffuse background
emission in the InfraRed Astronomical Satellite (IRAS) 60 and 100$\,\mu$m
maps and termed it `infrared cirrus' due to its complex texture.  The cirrus
has been associated with thermal emission from dust grains in the diffuse
interstellar medium (Beichman 1987).  Combining the angular resolution of the
IRAS observations with the photometric accuracy of the COsmic Background
Explorer (COBE)/Diffuse InfraRed Background Experiment (DIRBE) data, SFD
derived all-sky maps of the dust column density and mean temperature.  One of
the primary uses of these maps has been to correct extragalactic sources for
extinction and reddening.  

The SFD maps show structure in the cirrus down to the smallest angular scales
resolved by IRAS ($\approx6'$).  Gautier et~al.\ (1992) described the
structure in the IRAS data in terms of its angular power spectrum,
\begin{equation}
P(\kappa)\propto\kappa^\alpha~~~,
\end{equation}
where $P$ is the Fourier power, $\kappa$ is the reciprocal of the angular
scale, and $\alpha\sim-3$ is the spectral index.  Similarly, Guhathakurta \&
Cutri (1994) examined both IRAS 100$\,\mu$m maps and optical CCD surface
photometry of reprocessed starlight from dust grains: they found that power
on small scales was dominated by stars and galaxies, and since it was
impossible to completely separate out these components, they merely cleared
out the obvious compact ``objects'' (stars and galaxies) and smoothed the
residual image over arcminute scales; the resulting power spectrum had an
index $\alpha=-3$.  More recently, Kiss et~al.\ (2003) observed 13~fields
with Infrared Space Observatory (ISO)/ISOPHOT and found that the power
spectrum index varies from field to field over the range
$-5.3\le\alpha\le-2.1$ for angular scales larger than $3'$.  This work
extends the finding of $-3.6\le\alpha\le-0.5$ by Herbstmeier et~al.\ (1998)
in their earlier ISOPHOT-based study.  These last two~studies indicate that
$\alpha\sim-3.0$ is not universal. 
  
In this study, the scatter in the HB of the differentially-reddened globular
cluster NGC~4833 is used to measure the angular power spectrum of dust on
scales smaller than the angular resolution of the thermal emission maps
(e.g.,~IRAS, DIRBE, ISOPHOT).  For uncrowded data such as ours, the smallest
angular scale down to which this method can be applied is set by the surface
density of HB tracers which in turn determines the typical nearest-neighbor
separation; in crowding-limited situations, the smallest angular scale is a
few times the size of the stellar point spread function: few arcseconds for
ground-based images and sub-arcsecond for {\it Hubble Space Telescope\/}
images.

Because interstellar dust causes extinction and reddening of starlight and
reprocesses the energy, dust patchiness has a profound impact on the accuracy
of many astronomical measurements even at high Galactic latitudes.  Thermal
emission from cold dust can confuse CMBR anisotropy measurements.
Measurements of large-scale structure in galaxy surveys may be influenced by
spatial non-uniformities in the foreground dust.  Studies of stellar
populations are at risk because measurements of $T_{\rm eff}$ can be impacted
by reddening.  Extinction can affect the photometry of distance indicators
and thereby produce systematic errors in the distance-scale ladder.  It is
therefore important to quantify extinction/reddening variations on the
smallest scales possible.  While the SFD study does an excellent job of
characterizing moderate- to large-scale dust variations, their maps miss
power on scales smaller than a few~arcminutes.\\\\

The photometry of NGC~4833 is presented in \S\,2.  The construction of
simulated cluster data sets in discussed in \S\,3.  The $R_V$ measurement
methods are outlined in \S\,4, along with their application to NGC~4833 and
simulated data sets to estimate the accuracy of each method.  In \S\,5, we
describe the angular power spectrum of cirrus in the direction of NGC~4833.
Possible future extensions of this work are discussed in \S\,6.  The main
conclusions of this paper are summarized in \S\,7.

\section{The Data}

The globular cluster NGC~4833 is located near the equatorial plane of our
Galaxy ($l=304^\circ,~ b=-8^\circ$) in a dusty region of the constellation
Musca.  The cluster has a mean reddening of $\langle{E(B-V)}\rangle=0.32$~mag
and reddening variations of order $\pm0.05$~mag across the field (Melbourne
et~al.\ 2000).  Figure~1 of Melbourne et~al.\ shows an image of the dusty
features in the region around the cluster including the Coalsack nebula.

\subsection{Multiband CCD Photometry}

The Johnson-Cousins $BVI$ stellar photometry of NGC~4833 used in this study
are drawn from Melbourne et~al.\ (2000).  A detailed description of the data
set is given in that paper so only a brief discussion appears here.  The CCD
images of the cluster were obtained in 1996 using the 0.9-m telescope at the
Cerro Tololo Inter-American Observatory\footnote{The Cerro Tololo
Interamerican Observatory, National Optical Astronomy Observatories, is
operated by the Association of Universities for Research in Astronomy, Inc.\
under cooperative agreement with the National Science Foundation.}.  The
data reduction was carried out with the DAOPHOT~II (Stetson 1994) stellar
photometry package.

\begin{figure}
\includegraphics[scale=0.7]{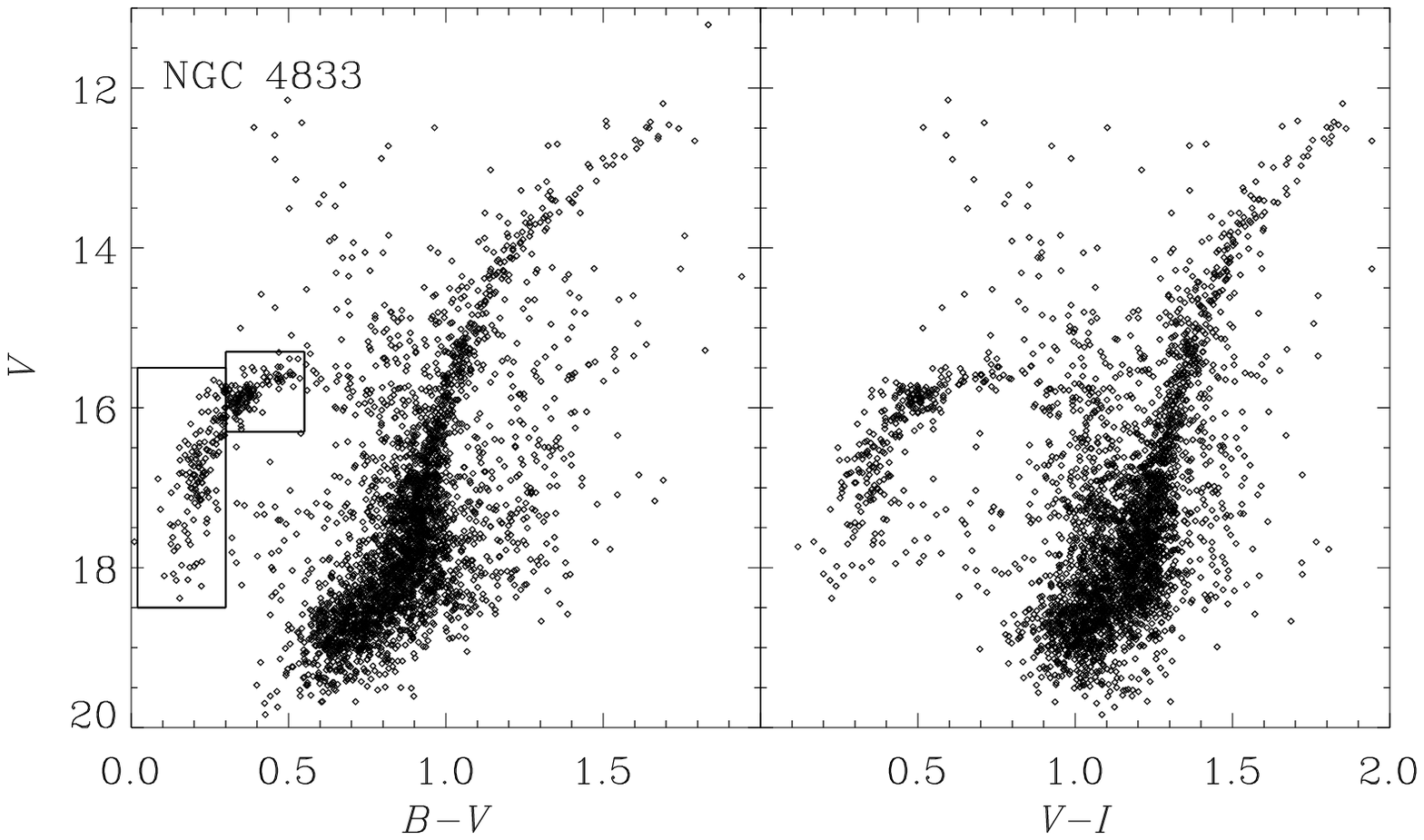}
\figcaption{\label{fig:fullCMD}
Color-magnitude diagrams in $V$ vs.\ $B-V$ (left) and $V$ vs.\ $V-I$ (right)
of a $13.5'\times13.5'$ field centered on the low-latitude Galactic globular
cluster NGC~4833.  Note the prominent curved blue horizontal branch.  The
boxes in the ($B-V$,~$V$) CMD are used to select 260~candidate cluster HB
stars; the analysis presented in this paper is based solely on this subset of
stars.
}
\end{figure}

Figure~\ref{fig:fullCMD} shows the ($B-V$,~$V$) and ($V-I$,~$V$) CMDs (left
and right panels, respectively) for the full $13.5'\times13.5'$ field.  Boxes
are drawn in the former CMD around the 260~objects that are considered to be
candidate cluster HB stars for the purposes of this study (possible non-HB
interlopers in the sample are discussed in \S\,4.1.1).  Extinction/reddening
variations across the field affect the width of the principal sequences of
the CMDs.  Figure~\ref{fig:radCMD} shows that the HB is relatively tight and
narrow for the central $r<2'$ region of the cluster, but increases in width
for stars distributed over a larger area in the outer region of the cluster.
This suggests that extinction/reddening variations across the field, rather
than photometric errors, are the cause of the large scatter in the HB; if
anything, photometric errors tend to be slightly worse in the inner region of
the cluster because of crowding.

A color-color diagram of cluster HB stars is plotted in the top panel of
Figure~\ref{fig:cc}, while the bottom panel shows the same after correcting
for reddening on a star-by-star basis.  The $E(B-V)$ value for each star is
taken from the SFD dust map, and an assumed $R_V$ value of~3.0 is plugged
into the CCM parameterization of the dust extinction law
(Eqn.~\ref{eqn:cardel}) to derive $R_I$ and the corresponding $E(V-I)$ value.
The reddening-corrected data (and the uncorrected data, for that matter) show
a linear relationship between $B-V$ and $V-I$.  A least-squares fit (solid
line) is made to the reddening-corrected data:
\begin{equation}
\label{eqn:cc}
(V-I)=-0.0068+1.27\,(B-V)~~~.
\end{equation}
This equation, in conjunction with the two-polynomial fiducial fit to the HB
in the ($B-V$,~$V$) CMD, is used to simulate $BVI$ photometry of HB stars
(\S\,3.2). 

\subsection{Photometric Errors}

Figure~\ref{fig:photerr} shows the typical $1\sigma$ photometric errors in
the $BVI$ bands for NGC~4833 stars as a function of their apparent visual
magnitude (Melbourne et~al.\ 2000).  The photometric errors, estimated using
the DAOPHOT~II software package (Stetson 1994), range from $\sigma_{\rm
phot}\sim0.01$~mag at the bright end of the HB ($V\sim15.5$) to $\sigma_{\rm
phot}\sim0.05$~mag at the faint end ($V\sim18$), and contribute noticeably to
the width of the HB.  The characteristic photometric error is defined to be
that at $V=17$, $\rm\sigma_{phot}(NGC~4833)\sim0.03$~mag, and this is what is
referred to throughout the rest of this paper.

\section{Simulated Cluster Data Sets}

A key aspect of our study is the use of simulated cluster data sets to fine
tune and test the $R_V$ measurement methods (\S\,4).  These simulated data
sets are designed to mimic the general properties of the NGC~4833 data set as
closely as possible.  Since we are also interested in applying these $R_V$
measurement methods to a broad range of data, the main set of simulations are
constructed over a two-dimensional grid of parameter space: three {\it
input\/} $R_V$ values and three~levels of photometric precision.  At each of
these 9 ($3\times3$) grid points, 100~Monte-Carlo realizations are carried
out---i.e.,~a total of 900~simulated clusters are constructed each containing
260~HB stars.  The rest of this section describes the step-by-step procedure
used to construct each simulated data set.

\subsection{Choosing Extinction Law Parameters}

Three~input $R_V$ values, 2.0, 3.0 and 4.0, are chosen for the simulated
cluster data sets.  These mimic a range of extinction law properties for the
foreground dust and span the range of values observed in lines of sight with
moderate to light reddening (\S\,1.1).  The CCM parameterization of the
Galactic extinction laws:
\begin{equation}
\label{eqn:cardel}
R_I \equiv A_V/E(V-I) = \frac{1}{0.32 + 0.6239/R_V}.
\end{equation}
is adopted here.  For a given choice of $R_V$, the visual extinction $A_V$ is
translated to its corresponding $E(B-V)$ and $E(V-I)$ color excesses.

\subsection{Creating an Artificial Horizontal Branch}

Each simulated star is assigned $B$- and $V$-band photometry via a random
drawing from a fiducial HB curve in the CMD.  The fiducial is based on
stitching together two~second-order polynomials that are fit to the
extinction/reddening-corrected ($B-V$,~$V$) HB of NGC~4833, where the
$E(B-V)$ correction for each star is obtained from the SFD dust map and $R_V$
is assumed to be~3.0.  Examples of such two-polynomial fiducial HBs are shown
in Figures~\ref{fig:Av_method} and \ref{fig:schlegel_method}.  The
``SFD-corrected'' cluster HB used for the fiducial fit is similar to the one
shown in the left-middle panel of
Figure~\ref{fig:cmd6}, except for the fact that the one illustrated has had
the mean extinction/reddening reapplied to it.  When drawing 260~stars from
the fiducial, the run of stellar density along the simulated HB is made to
mimic that observed along the cluster HB.  This is done by dividing the
fiducial into sections and demanding that the number of stars in each section
match that in the corresponding section of the cluster HB.  Simulated
$I$-band photometry is obtained from the linear fit to the ($B-V$,~$V-I$)
color-color diagram of cluster HB stars (Fig.~\ref{fig:cc},
Eqn.~\ref{eqn:cc}).

\subsection{Modeling Reddening Variations Across the Cluster}

The width of the HB can be used to quantify the effect of
extinction/reddening variations.  Figure~\ref{fig:HBscat} shows the RMS
scatter of HB stars in NGC~4833 as a function of their $V$ magnitude (solid
line).  The RMS scatter is calculated based on the {\it shortest distance\/}
of each HB star from the fiducial curve (i.e.,~distance to the nearest
tangent point on the curve).  The fiducial consists of two~second-order
polynomials fit to HB stars within $2'$ of the cluster center.  The dotted
line in Figure~\ref{fig:HBscat} is a model prediction for the HB scatter that
would result from photometric errors alone, the dashed line shows the effect
of large-scale extinction/reddening variations from SFD plus photometric
error, and the dot-dashed line the effect of small- and large-scale
extinction/reddening variations plus photometric error.  It is clear from the
scatter at the bright end of the HB ($V\lesssim17.5$) that only the
dot-dashed line matches the NGC~4833 data.  The following subsections
describe in turn the simulations behind the three~model lines.

\subsubsection{In the Absence of Reddening Variations}

A special set of 100~photometric-error-only (``$\sigma_{\rm phot}$-only'')
simulations containing {\it no\/} reddening variations is constructed for
illustration purposes only; this set is not used in any of the $R_V$
determination exercises described in \S\,4.  Each Monte Carlo realization in
this special set involves three~steps: ({\bf 1})~An artificial HB is
constructed as described in \S\,3.2 above; ({\bf 2})~The mean extinction and
reddening of the cluster, $\langle{A_V}\rangle=1.0$ and
$\langle{E(B-V)}\rangle=0.32$ (based on $R_V=3.0$), are applied to all
simulated stars to bring them into the same range of apparent $BVI$
magnitudes as the NGC~4833 HB stars; and ({\bf 3})~Gaussian photometric
errors are added assuming $\rm\sigma_{phot}(sim)=\sigma_{phot}(NGC~4833)$.
The details of step~(3) are given in \S\,3.4 below.

The dotted line in Figure~\ref{fig:HBscat} is the scatter measured in
$100\times260$~simulated HB stars.  The mismatch with the NGC~4833 data
(solid line) demonstrates that photometric errors alone cannot explain the
width of the cluster's HB and there must be additional sources of scatter in
the HB.  (The dotted line is much smoother than the solid line because it is
based on 100$\times$ as many stars.)

\subsubsection{Large-Scale Variations}

The obvious additional source of HB scatter is differential
extinction/reddening across the cluster.  Large-scale variations based on the
SFD map are explored in this section.  Their maps provide $E(B-V)$ color
excesses as a function of sky position, but have an angular resolution of
$\sim6'$ and therefore contain power only on scales larger than this.  The
SFD map includes NGC~4833 and the region around it, and indicates that the
mean reddening along this line of sight is
$\langle{E(B-V)}\rangle\sim0.3$~mag with variations on the order of
$\pm0.05$~mag.

Following the procedure described above for the ``$\sigma_{\rm phot}$-only''
simulations (\S\,3.3.1), another special set of 100~simulations is
constructed containing only large-scale extinction/reddening variations and
photometric errors.  Again, these special ``SFD+$\sigma_{\rm phot}$''
simulations are created for illustrative purposes alone.  The only difference
is in step~(2) of the procedure: instead of applying the same (mean)
extinction and reddening values to all simulated HB stars, each star is
randomly assigned the sky position of one of the 260~HB stars in NGC~4833 and
the corresponding SFD $E(B-V)$ value, and associated $A_V$ and $E(V-I)$
values based on $R_V=3.0$, are applied to the $BVI$ magnitudes of the
simulated star.  By assigning actual sky positions to the simulated HB stars,
we ensure that they mimic the spatial distribution of NGC~4833's HB stars and
range of large-scale extinction/reddening variations across the cluster data
set.

The dashed line in Figure~\ref{fig:HBscat}, based on $100\times260$~simulated
HB stars, shows the scatter resulting from a realistic amount of large-scale
extinction/reddening variations and photometric error.  Even this line falls
a little short of explaining the HB scatter observed in the cluster (solid
line).

\subsubsection{Small-Scale Variations}

The most probable cause of this shortfall is the coarse resolution of the SFD
map which averages over structure in the dust distribution on angular scales
smaller than $\sim6'$.  Small-scale variations are incorporated into the
simulated data sets by adding random Gaussian fluctuations with an amplitude
proportional to the SFD reddening value at that location:
\begin{equation}
\label{eqn:smallscale}
E(B-V)_{\rm LS+SS}=E(B-V)^{\rm SFD}[1+f_i^{\rm SS}]
\end{equation}
where $f_i^{\rm SS}$ is a random number (for the $i$-th star) drawn from a
Gaussian distribution with zero mean and width $\sigma^{\rm SS}=0.06$.

This brings us back to the main set of simulations used for testing the $R_V$
determination methods in \S\,4.  They are constructed using a procedure
similar to that used for the special set of ``SFD+$\sigma_{\rm phot}$''
simulations (\S\,3.3.2) but with two changes.  First, the SFD reddening
values in step~(2) of the procedure are replaced by $E(B-V)_{\rm LS+SS}$ from
Eqn.~\ref{eqn:smallscale} above.  Second, a range of $R_V$ values (\S\,3.1)
and photometric error levels (\S\,3.4) are used.

With the above choice of Gaussian width for the small-scale dust variations,
$\sigma^{\rm SS}=0.06$, and for
$\rm\sigma_{phot}(sim)=\sigma_{phot}(NGC~4833)$ and $R_V=3.0$, the scatter in
the simulated HB approaches that of NGC~4833's HB (dot-dashed vs.\ solid
lines in Fig.~\ref{fig:HBscat}).  This is a reasonable estimate of the amount
of smoothing of dust fine-scale structure in the SFD maps (\S\,5.1;
Fig.~\ref{fig:powspec}), and is well within the 10\% uncertainty in reddening
values estimated by the authors.

\subsection{Accounting for Photometric Error}

The final step in the construction of the main set of simulated HB data sets
is the inclusion of photometric error.  Since the photometric uncertainty is
a function of stellar brightness, this step must be carried out after the
inclusion of (large- and small-scale) dust effects so that each star is at
the appropriate apparent brightness.  Gaussian random noise is added to the
$BVI$ magnitudes of each star.  Three~levels of photometric error are
simulated: (1)~$\rm\sigma_{phot}(sim)=\sigma_{phot}(NGC~4833)$, with the
latter being the DAOPHOT-based errors shown in Figure~\ref{fig:photerr};
(2)~$\rm\sigma_{phot}(sim)=0.5\,\sigma_{phot}(NGC~4833)$; and
(3)~$\rm\sigma_{phot}(sim)=0.1\,\sigma_{phot}(NGC~4833)$.

\section{Measuring $R_V$}

We employ three~methods for measuring $R_V$ in the direction of 
differentially-reddened globular clusters.  Each method relies on the premise 
that the width of the blue HB will be minimized when the appropriate $R_V$
value is used to correct for extinction and reddening.  A description of each
method and its application to the NGC~4833 and simulated cluster data sets
follows.
  
\subsection{$A_V$~RMS Method}

\subsubsection{Description of Technique}

The ``$A_V$~RMS'' method relies on three-color photometry of
differentially-reddened globular clusters.  Figure~\ref{fig:Av_method} shows
the ($B-V$,~$V$) and ($V-I$,~$V$) CMDs of a simulated HB (left and right
panels, respectively).  A fiducial, consisting of two~second-order
polynomials stitched together (solid curve; \S\,3.2), is fit to the HB stars
within $2'$ of the cluster center where differential extinction/reddening is
relatively small; this is done independently for the ($B-V$,~$V$) and
($V-I$,~$V$) CMDs.  The relative visual extinction of each star is calculated
to be the vertical distance between the star and the point where the
reddening vector drawn from the star intersects the fiducial
(Fig.~\ref{fig:Av_method} inset); it is designated $(\delta{A_V})^{BV}$ and
$(\delta{A_V})^{VI}$ for the ($B-V$,~$V$) and ($V-I$,~$V$) CMDs,
respectively.\footnote{Throughout this paper, the symbol ``$\delta{A_V}$'' is
used to denote the relative visual extinction of a star based on a single
fiducial/CMD, whereas ``$\Delta{A_V}$'' is used to denote the {\it
difference\/} between two $\delta{A_V}$ values, say between two~CMDs for a
given star or between a pair of stars.}  The $(\delta{A_V})^{BV}$ measurement
is based on an assumed/test value of $R_V$, the slope of the reddening vector
in the ($B-V$,~$V$) CMD.  For the $(\delta{A_V})^{VI}$ calculation, the CCM
parameterization given in Eqn.~\ref{eqn:cardel} is used to translate $R_V$ to
$R_I$, the slope in the ($V-I$,~$V$) CMD.

The difference between the two $\delta{A_V}$ values is defined to be 
$(\Delta{A_V})^{BV,VI}\equiv\,(\delta{A_V})^{BV}-(\delta{A_V})^{VI}$.  For a
given test value of $R_V$, the RMS scatter of $(\Delta{A_V})^{BV,VI}$ is
computed over all HB stars.  This is done in two~iterations, with 3-$\sigma$
rejection of outliers after the first iteration to eliminate any non-HB stars
in the sample (e.g.,~cluster blue stragglers or Galactic field main sequence
turnoff stars in the foreground---encircled symbols in upper panels of
Fig.~\ref{fig:radCMD}).  Non-HB interlopers are expected to have large
deviations from the HB fiducial with effective $\delta{A_V}$ values that are
different between the two~CMDs.

The calculation of the RMS of $(\Delta{A_V})^{BV,VI}$ is repeated for a
series of test $R_V$ values between 0.5 and 5.5 in steps of~0.25.  The
$\delta{A_V}$ parameters from the two~CMDs are expected to approach each
other when the test value approaches the true $R_V$ of the dust in the
foreground of the cluster.  In other words, the RMS of
$(\Delta{A_V})^{BV,VI}$ should go through a minimum at the correct $R_V$
value.  A polynomial fit is made to the RMS of $(\Delta{A_V})^{BV,VI}$ vs.\
test $R_V$ data points to interpolate to the minimum.

\subsubsection{Application to Data and Simulations}

Figure~\ref{fig:Av_results} shows the RMS scatter in $(\Delta{A_V})^{BV,VI}$
vs.\ test $R_V$ value, the result of applying the $A_V$~RMS method to
NGC~4833 (filled diamonds) and three~sample simulated cluster data sets (open
triangles).  The simulated data sets are all based on an input $R_V$ value of
3.0, but have different levels of photometric error: 10\%, 50\%, and 100\%
that of the NGC~4833 data set (\S\,3.4), with the RMS sequences arranged in
order of $\rm\sigma_{phot}(sim)$ increasing upwards.  As expected, each
sequence of RMS of $(\Delta{A_V})^{BV,VI}$ vs.\ test $R_V$ goes through a
clear minimum (large open square) near $R_V=3.0$.  The depth/sharpness of the
RMS minimum decreases with increasing photometric error and the power of the
method diminishes.

The results for all $9\times100$~simulated cluster data sets are summarized
in Table~\ref{tbl:sim}.  For
$\rm\sigma_{phot}(sim)=0.1\,\sigma_{phot}(NGC~4833)$, the $A_V$~RMS method
reproduces the correct $R_V$ to within $\pm0.1$, as measured by the standard
deviation of measured (output) $R_V$ values for 100~simulated data sets.  The
scatter in output $R_V$ increases to $\pm0.7$ for
$\rm\sigma_{phot}(sim)=\sigma_{phot}(NGC~4833)$.  The (absolute) precision of
the method appears to decrease with increasing input $R_V$, again judging
from the scatter of output $R_V$ values, although the fractional error
$\sigma(R_V)/R_V$ remains roughly constant.  It is important to note that the
output $R_V$ from this method tends to be biased low for moderate to large
photometric errors and input $R_V\gtrsim3$.  For example, simulated data sets
with input $R_V=3.0$ and $\rm\sigma_{phot}(sim)=\sigma_{phot}(NGC~4833)$ (a
good match to the real cluster data) have a mean output
$\langle{R_V}\rangle=2.60\pm0.7$.  This bias of $-0.4$ in
$\langle{R_V}\rangle$ is significant compared to the error in the mean,
$0.7/\sqrt{100}=0.07$, and must be taken into account in the determination of
$R_V$ for an actual cluster data set.

The cause of the bias towards low $R_V$ may be explained as follows.
Photometric error produces scatter in the HB which, unlike the scatter caused
by variable extinction/reddening, is {\it uncorrelated\/} between the
two~CMDs---i.e.,~the resulting values of $(\delta{A_V})^{BV}$ and
$(\delta{A_V})^{VI}$ are different.  These errors are particularly important
at faint magnitudes where the HB is vertical and the effect is a spread in
color.  All else being equal, choosing a smaller $R_V$ (and $R_I$) translates
these color errors into smaller $\delta{A_V}$ values and hence to smaller
($\Delta{A_V}$)$^{BV,VI}$ differences as well.  This biases the RMS minimum
towards lower $R_V$ values.

The RMS of $(\Delta{A_V})^{BV,VI}$ for the NGC~4833 data set displays a
significant minimum at $R_V=2.5\pm0.7$ (open square associated with filled
diamonds in Fig.~\ref{fig:Av_results}), where the error bar is taken directly
from the scatter in output $R_V$ for comparable simulated data sets.
Correcting for the bias in the measured $\langle{R_V}\rangle$ for such
simulations, our best guess for the $R_V$ towards NGC~4833 is $2.9\pm0.7$.

\subsection{$A_V$~Slope Method}

\subsubsection{Description of Technique}

The ``$A_V$~Slope'' method is closely related to the $A_V$~RMS method in that
it relies on the same $(\delta{A_V})^{BV}$ and $(\delta{A_V})^{VI}$ values
determined for each star in \S\,4.1.1 above.  Instead of computing the
difference between the two $\delta{A_V}$ values as in the $A_V$~RMS method,
one is plotted against the other and a linear fit made.  This is illustrated
in Figure~\ref{fig:Av_slope_method} for a simulated cluster data set created
with (input) $R_V=3.0$; four~test $R_V$ values are shown.  The linear fit in
each panel is effectively a bivariate one: one of the dashed lines is a fit
treating the $x$- and $y$-axes as the independent and dependent variables,
respectively, and vice versa for the other dashed line; their slopes are
averaged to get the solid line.  As expected, the slope of the linear fit
approaches unity when the test $R_V$ approaches the true (input) $R_V$ value.
As in \S\,4.1.1, a polynomial fit is made to the slope vs.\ test $R_V$ data
to interpolate to unit slope.

\subsubsection{Application to Data and Simulations}

Figure~\ref{fig:slope_results} shows results from the application of the
$A_V$~Slope method: the slope of the relation between $(\delta{A_V})^{BV}$
and $(\delta{A_V})^{VI}$ is plotted versus test $R_V$.  The format of this
figure is the same as for Figure~\ref{fig:Av_results}, except that a slope of
unity (rather than an RMS minimum) corresponds to the best-fit $R_V$ value.
The $A_V$~Slope method results for NGC~4833 and simulated data sets are
summarized in Table~\ref{tbl:sim}.

Applying this method to $9\times100$~simulated data sets, we learn that, like
the $A_V$~RMS method, the power of the $A_V$~Slope method decreases with
increasing photometric errors.  The scatter in output $R_V$ is $\pm0.1$ for
$\rm\sigma_{phot}(sim)=0.1\,\sigma_{phot}(NGC~4833)$, but increases to 
$\pm0.6$ for $\rm\sigma_{phot}(sim)=\sigma_{phot}(NGC~4833)$.  Some of the
9~simulation categories display a small but significant bias (positive or
negative) in the measured (output) $\langle{R_V}\rangle$, but not the
category directly relevant to NGC~4833.  The cause of this bias is not as
readily apparent as for the ``$A_V$~RMS'' method (\S\,4.1.2).  For the
NGC~4833 data set, the $A_V$~Slope method gives $R_V=2.6\pm0.6$.

\subsection{Optical/IR Method}

\subsubsection{Description of Technique}

The ``Optical/IR'' method is different from the previous two~methods in that
it relies on two-color photometry (i.e.,~a single CMD which, in our case, is
$V$ vs.\ $B-V$) and the SFD reddening map.  Using a test $R_V$ value, $A_V$
values are derived for all HB stars by scaling the corresponding SFD $E(B-V)$
values.  These color and magnitude corrections are applied on a star-by-star
basis.  Figure~\ref{fig:schlegel_method} shows the ($B-V$,~$V$) CMD of a
simulated HB before (left) and after (right) making the extinction/reddening
corrections.  The latter HB is slightly tighter, but not dramatically so
because the SFD reddening values smooth over small-scale dust variations.
The RMS scatter of the corrected HB stars is computed with respect to a
best-fit fiducial curve (two~second-order polynomials---solid line in
Fig.~\ref{fig:schlegel_method}) based on ``shortest distance'' (see \S\,3.3).
This process is repeated for a range of test $R_V$ values as for the other
two~methods.  The RMS scatter is expected to go through a minimum when the
test $R_V$ approaches the true $R_V$ value.  As in \S\,4.1.1, a polynomial
fit is made to the HB scatter vs.\ test $R_V$ data to interpolate to the
minimum.

\subsubsection{Application to Data and Simulations}

Figure~\ref{fig:schlegel_results} shows results from the application of the
Optical/IR method: RMS scatter of HB stars versus test $R_V$.  The format of
this figure is the same as for Figure~\ref{fig:Av_results}.  The Optical/IR
results for NGC~4833 and simulated data sets are summarized in
Table~\ref{tbl:sim}.

Tests of the method on $9\times100$~simulated cluster data sets indicate that
the method has a low level of predictive power ($\sim\pm0.6$ in $R_V$) over
the full range of photometric errors simulated.  This is because the main
source of error here is uncertainty in the SFD reddening values, including
small-scale dust variations that get smoothed out by the relatively coarse
resolution of their map.  The Optical/IR method gives $R_V=3.5\pm0.7$ for
NGC~4833.

\subsection{Discussion}

Our best estimate of the extinction law slope in the direction of NGC~4833 is
$R_V=3.0\pm0.4$, based on a weighted average of the values from the
three~determination methods (see Table~\ref{tbl:sim}).  Before averaging, the
output $R_V$ value from the $A_V$~RMS method is corrected for measurement
bias: $2.5\rightarrow2.9$ (\S\,4.1.2).  The weights are proportional to
$1/[\sigma(R_V)]^2$ where $\sigma(R_V)$ is the uncertainty in output $R_V$
for a given determination method as derived from the simulations.

Figure~\ref{fig:cmd6} illustrates the effectiveness of the star-by-star
extinction/reddening corrections based on the $R_V$ determination methods
described earlier in this section.  The top two~panels show the uncorrected
NGC~4833 HB in the ($B-V$,~$V$) and ($V-I$,~$V$) CMDs (left and right panels,
respectively).  The photometry in the middle two~panels has been corrected
for extinction/reddening based on the SFD $E(B-V)$ maps, using a best-fit
$R_V=3.0$ and the corresponding $R_I$ from Eqn.~\ref{eqn:cardel} to translate
the $B-V$ color excess to $A_V$ and $E(V-I)$.  Note, the mean extinction and
reddening of the cluster have been added back to the data for ease of
comparison with the top panels.  The HB is somewhat tighter after the
SFD-based correction but the effect is fairly subtle, a result of missing
small-scale dust structure in the SFD map and possibly other sources of
uncertainty as well.

The correction in the bottom two~panels of Figure~\ref{fig:cmd6} is based on
information derived from $BVI$ photometry of the cluster HB.  The relative
visual extinction measures derived from the two~CMDs, $(\delta{A_V})^{BV}$
and $(\delta{A_V})^{VI}$ (\S\,4.1.1), are averaged to obtain a 
$\langle\delta{A_V}\rangle$ value for each HB star.  This is translated to
color excesses in the two~CMDs using the best-fit $R_V=3.0$ and its
corresponding $R_I$ (Eqn.~\ref{eqn:cardel}).  The dramatic tightening of the
HB indicates that the scatter is strongly correlated between ($B-V$,~$V$) and
($V-I$,~$V$) CMDs---e.g.,~if a star is displaced to the right of/below the HB
fiducial in one CMD, it tends to be displaced in the same direction and by
the same amount (or by a proportional amount in color) in the other CMD.  The
most natural explanation of this correlated HB scatter is of course
differential extinction/reddening in the NGC~4833 data set.

\begin{figure}
\includegraphics[scale=0.7]{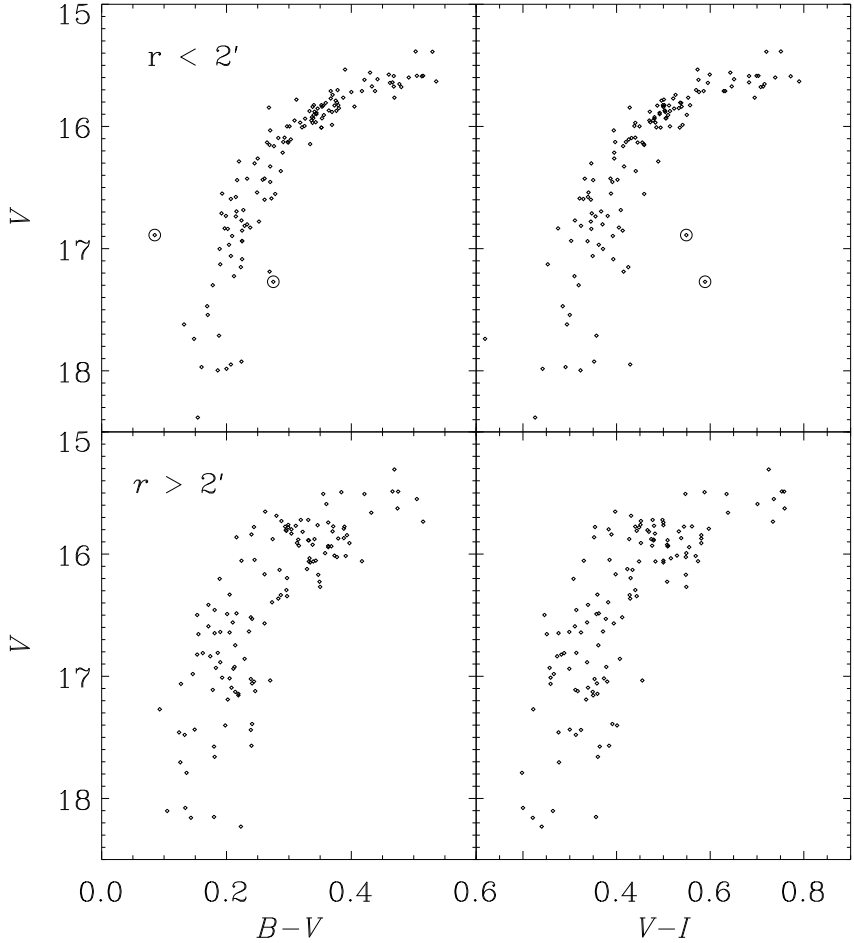}
\figcaption{\label{fig:radCMD}
Same as Figure~\ref{fig:fullCMD}, except only candidate HB stars are shown,
divided into two~groups based on their projected distance from the cluster
center: $r<2'$ (upper panels) and $r>2'$ (lower panels).  The HBs are tighter
in the central CMDs because they are based on a relatively small area of the
sky across which there is not much differential extinction/reddening.  By
contrast, the HBs in the lower panels are significantly broadened by the
effects of differential extinction/reddening; as a result, the scatter
`pattern' is similar between left and right CMDs.  The two~encircled dots in
the upper panels may be non-HB stars (\S\,4.1.1).
}
\end{figure}

\begin{figure}
\includegraphics[scale=0.7]{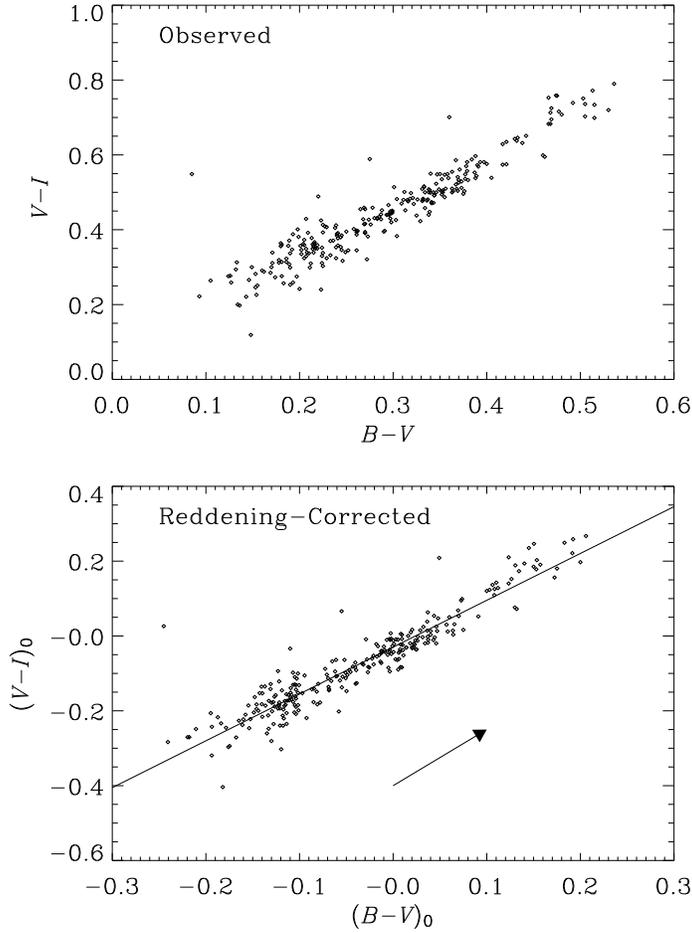}
\figcaption{\label{fig:cc}
The $B-V$ versus $V-I$ color-color diagram of NGC~4833 before (top) and after
(bottom) correcting for differential reddening.  The correction is taken from
the Schlegel et~al.\ (1998) dust maps with $R_V=3.0$.  The solid line is a
least-squares linear fit to the reddening-corrected data; it is used in the
construction of artificial HBs for simulated cluster data sets.  The arrow
indicates the direction of the reddening vector.  The reddening-corrected
sequence is only marginally tighter than the uncorrected one; the change is
subtle because the SFD map averages over small-scale dust variations and the
reddening vector runs nearly parallel to the sequence.
}
\end{figure}

\begin{figure}
\includegraphics[scale=0.7]{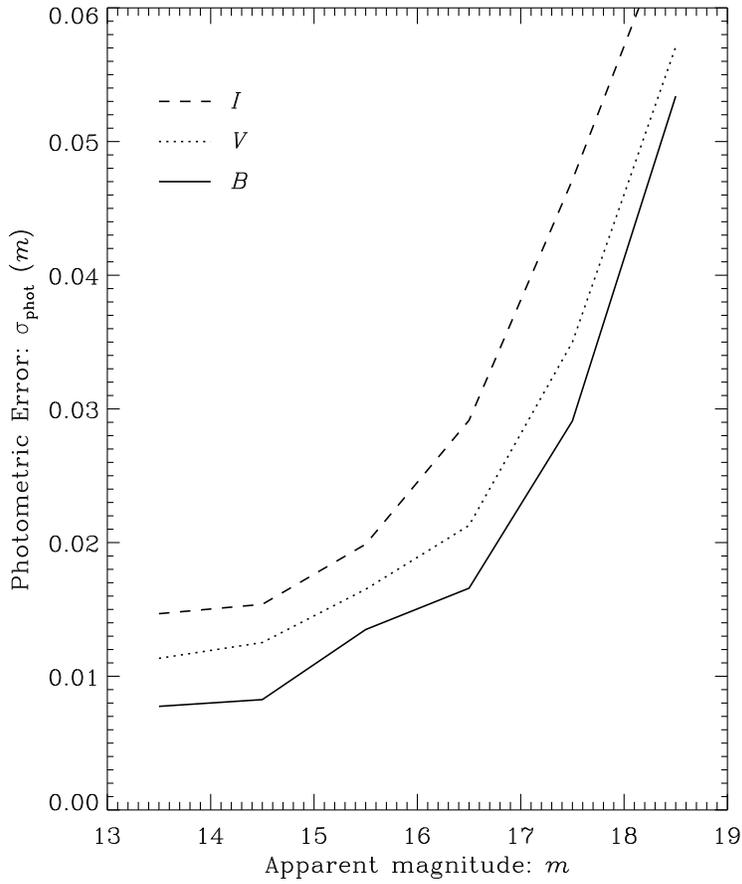}
\figcaption{\label{fig:photerr}
The mean $1\sigma$ photometric errors of HB stars in the NGC~4833 data set
are plotted as a function of apparent magnitude $m$ in the $B$, $V$, and $I$
bands.  The errors were derived from the DAOPHOT~II (Stetson 1994)
PSF-fitting routines as described in Melbourne et~al.\ (2000).
}
\end{figure}

\begin{figure}
\includegraphics[scale=0.7]{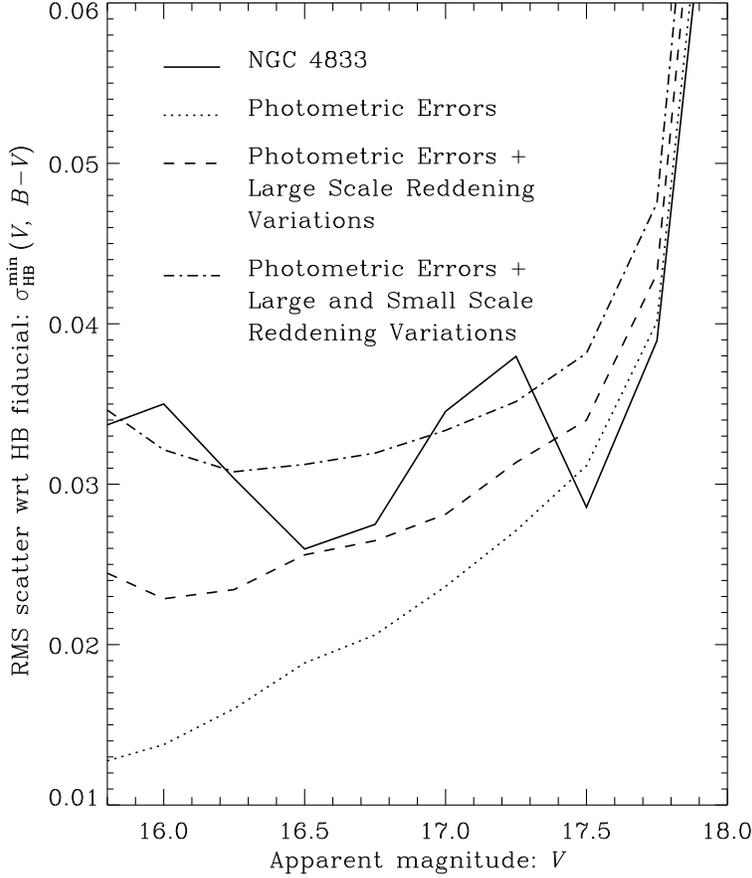}
\figcaption{\label{fig:HBscat}
The RMS scatter of NGC~4833's HB stars, based on ``shortest distance'' from a
fiducial line in the ($B-V$,~$V$) CMD (\S\,3.3), as a function of $V$
magnitude (solid line).  Each smooth curve represents 100~simulated cluster
data sets.  The dotted line is based on the special ``$\sigma_{\rm
phot}$-only'' simulations (\S\,3.3.1) constructed with mean
extinction/reddening and photometric errors similar to those of NGC~4833, but
with {\it no\/} spatial reddening variations.  The dashed line is based on
the special ``SFD+$\sigma_{\rm phot}$'' simulations (\S\,3.3.2) that also
include large-scale extinction/reddening variations from the Schlegel et~al.\
(1998) map.  The dot-dashed line is based on simulations that include large-
and small-scale extinction/reddening variations and photometric error
(\S\,3.3.3); all three~factors are needed to explain the scatter seen in the
bright portion ($V\lesssim17.5$) of the cluster HB.
}
\end{figure}

\begin{figure}
\includegraphics[scale=0.7]{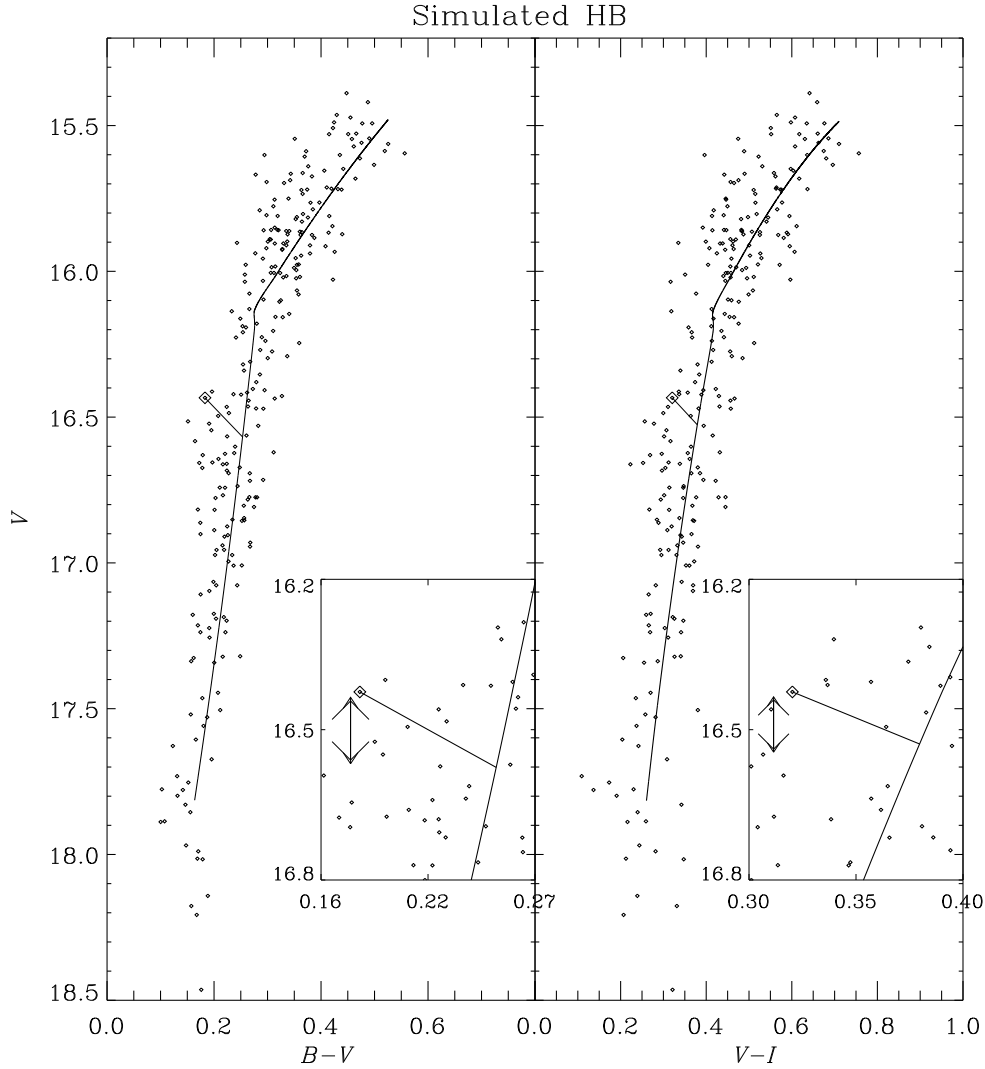}
\figcaption[Av_method.ps]{\label{fig:Av_method}
Color-magnitude diagrams of a simulated cluster horizontal branch for $B-V$
(left) and $V-I$ (right) color baselines.  The simulated data set includes
the effects of photometric error and large- and small-scale
extinction/reddening variations (\S\,3).  The fiducial HB in each panel
(solid line) is a fit of two~second-order polynomials to stars within the
central $r<2'$ of the simulated cluster (where differential
extinction/reddening are relatively small).  A sample reddening vector is
shown in each panel from a star to the fiducial HB.  The region around this
vector is magnified in the inset, and an arrow indicates the relative visual
extinction [$(\delta{A_V})^{BV}$ and $(\delta{A_V})^{VI}$] of the HB star
with respect to the fiducial (\S\,4.1.1).
}
\end{figure}

\begin{figure}
\includegraphics[scale=0.7]{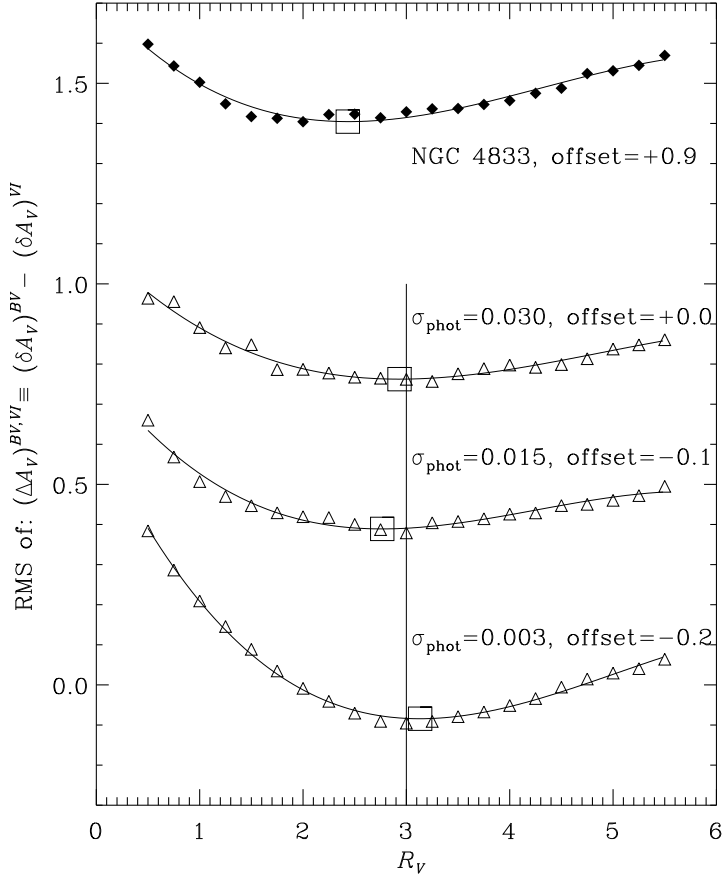}
\figcaption[RMSAv.ps]{\label{fig:Av_results}
Results from the application of the ``$A_V$~RMS'' method to NGC~4833 (filled
diamonds) and three~simulated cluster data sets (open triangles), each with a
different level of photometric error.  The RMS scatter in
$(\Delta{A_V})^{BV,VI}$, the difference between relative $A_V$ estimates from
the ($B-V$,~$V$) and ($V-I$,~$V$) CMDs (see \S\,4.1), is plotted for a range
of test $R_V$ values.  The input $R_V$ value used to construct all of these 
simulations is 3.0 (vertical line).  The large open square represents the
measured (output) $R_V$ value in each case, marked by the RMS minimum which
is interpolated using a polynomial fit (solid curve); the minimum is sharpest
for the simulation with the smallest photometric error.  Note, the sequences
have been offset vertically from one another by the amount indicated for the
sake of clarity.
}
\end{figure}

\begin{figure}
\includegraphics[scale=0.7]{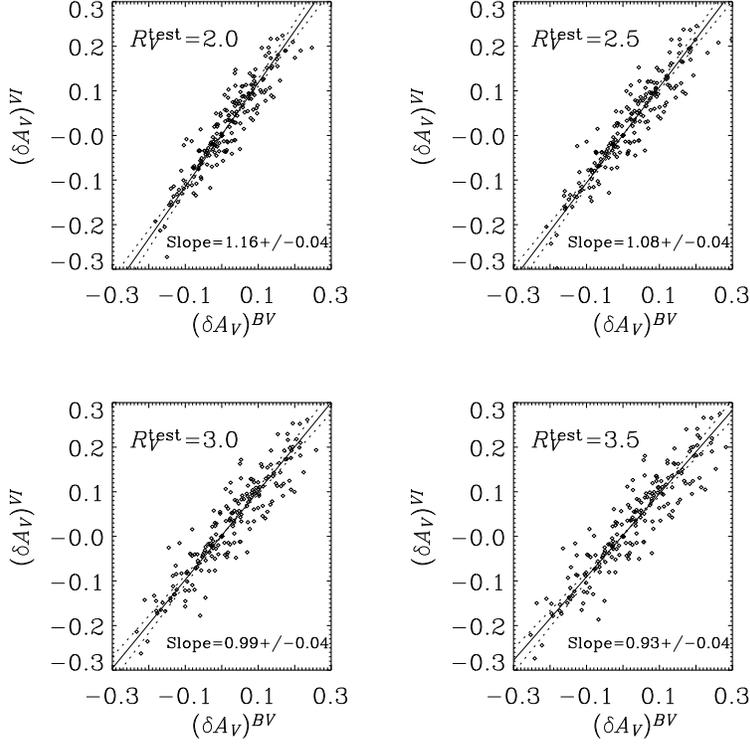}
\figcaption[slope_method.ps]{\label{fig:Av_slope_method}
Star-by-star comparison of the $(\delta{A_V})^{BV}$ and $(\delta{A_V})^{VI}$
values measured from the CMDs of a simulated cluster data set designed to
mimic the NGC~4833 data set (\S\,4.2.1).  Each of the four~panels assumes a
different test $R_V$ value for the purposes of measuring the $\delta{A_V}$
parameters.  The relationship between the two~$\delta{A_V}$ measurements
tends to be linear.  The dashed lines show two~linear fits to the data: one
uses the $x$ axis as the independent variable and the $y$ axis as the
dependent one, and vice versa for the other fit.  The solid line represents
the mean of these two~fits, with the best-fit slope indicated.  The slope of
the solid line approaches unity when the test $R_V$ equals the input
$R_V=3.0$ used to construct the simulated cluster data set.
}
\end{figure}

\begin{figure}
\includegraphics[scale=0.7]{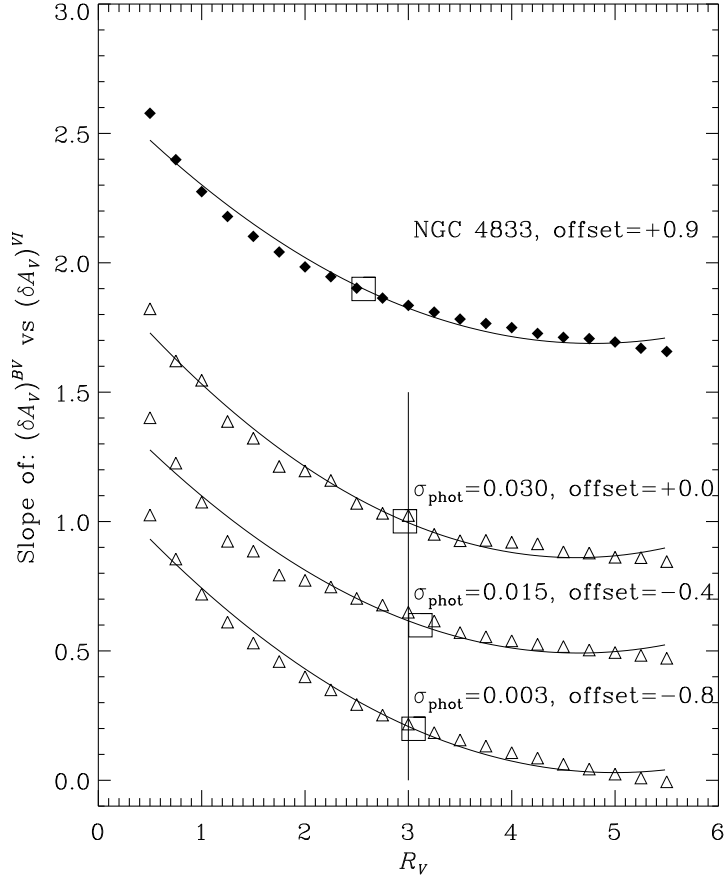}
\figcaption[slope_method.ps]{\label{fig:slope_results}
Same as Figure~\ref{fig:Av_results} for the ``$A_V$~Slope'' method (\S\,4.2),
with the $y$~axis showing the slope of the $(\delta{A_V})^{BV}$ versus
$(\delta{A_V})^{VI}$ relation (solid lines in
Fig.~\ref{fig:Av_slope_method}).  A slope of unity, interpolated via a
polynomial fit (solid curve), corresponds to the best-fit output $R_V$ value
(large open square).  The slope vs.\ test $R_V$ sequence is steepest for the
simulation with the smallest photometric error, indicating greatest
discriminating power for $R_V$ determination.
}
\end{figure}

\begin{figure}
\includegraphics[scale=0.7]{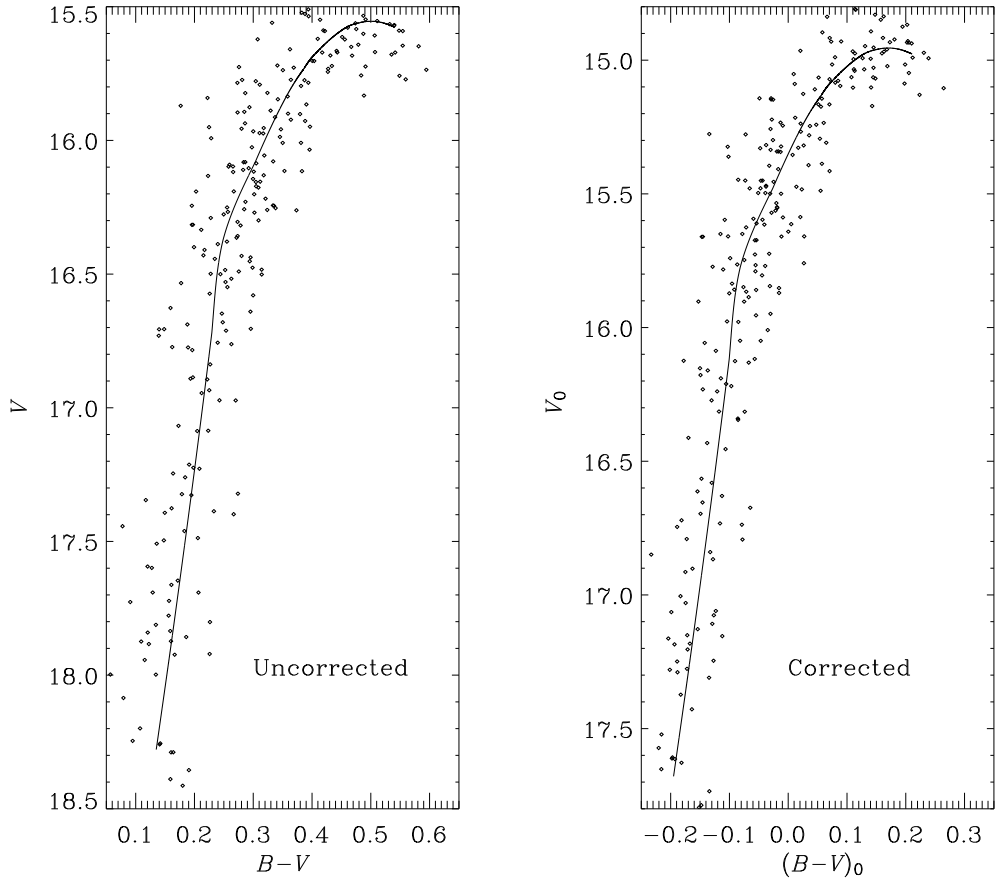}
\figcaption[shlegal_mneathod.ps]{\label{fig:schlegel_method}
(Left)~The ($B-V$,~$V$) color-magnitude diagram of a simulated cluster
horizontal branch, including the effects of photometric error and large- and
small-scale extinction/reddening variations.~~ (Right)~Same, after correcting
each star for extinction/reddening using the Schlegel et~al.\ (1998)
reddening map.  The correction has limited success in tightening the HB
because the SFD map cannot resolve small-scale dust variations.  The HB
fiducial (solid curve) in the right panel is a fit to the corrected data of
two~second-order polynomials stitched together; the fiducial in the left
panel is the same, but with the mean extinction/reddening added to match the
uncorrected data.
}
\end{figure}

\begin{figure}
\includegraphics[scale=0.7]{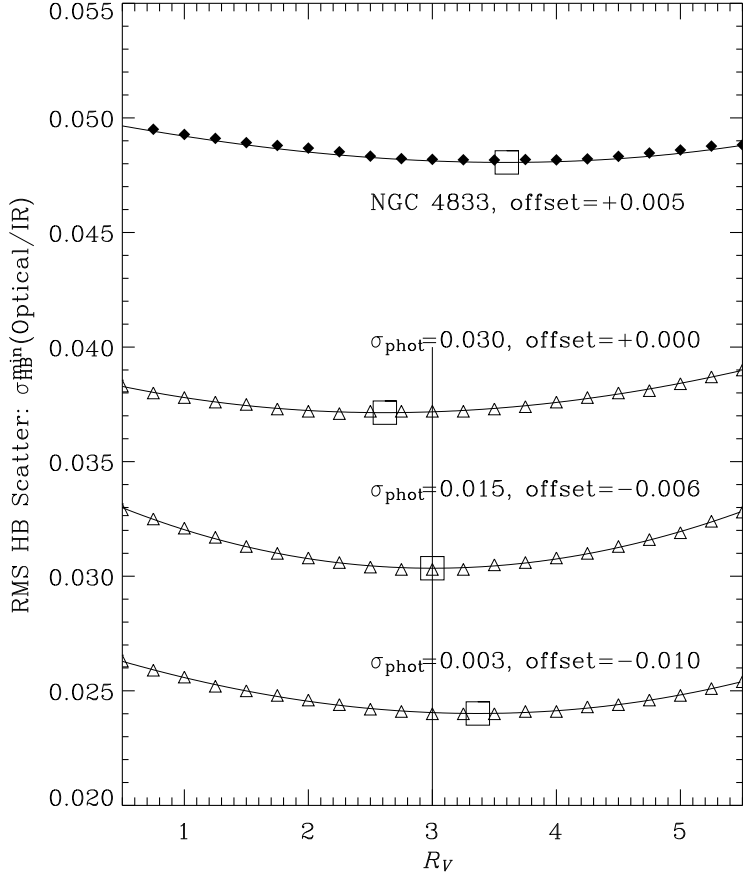}
\figcaption[RMSschlegal.ps]{ \label{fig:schlegel_results}
Same as Figure~\ref{fig:Av_results} for the ``Optical/IR'' method (\S\,4.3),
with the $y$~axis showing the RMS scatter of extinction/reddening-corrected
HB stars in a ($B-V$,~$V$) CMD.  The scatter is computed based on ``shortest
distance'' from a polynomial fiducial line (see \S\,3.3).  The $E(B-V)$
reddening correction is taken from the Schlegel et~al.\ (1998) dust map; it
is converted to $A_V$ using a test $R_V$ value.  This method has relatively
low discriminating power for $R_V$ determination because the SFD map averages
over small-scale reddening variations.
}
\end{figure}

\begin{figure}
\includegraphics[scale=0.7]{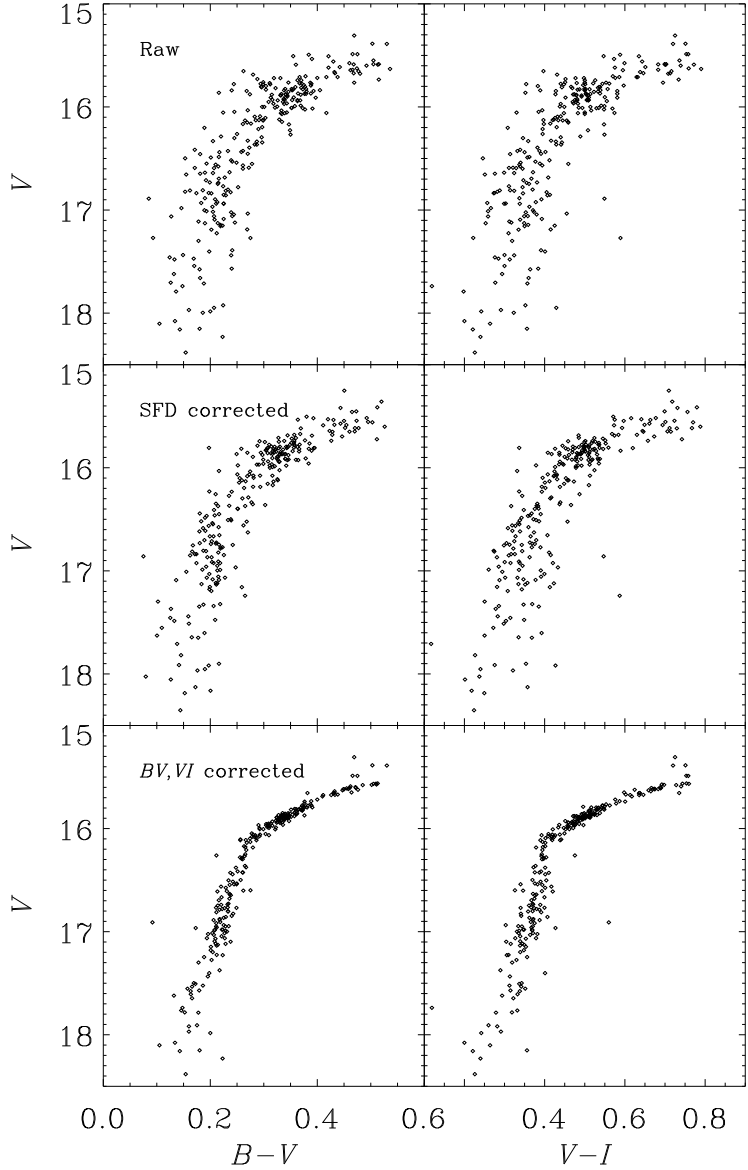}
\figcaption[cmd6.ps]{\label{fig:cmd6}
Color-magnitude diagrams showing the horizontal branch of NGC~4833 for $B-V$
(left) and $V-I$ (right) color baselines.  The top panels show the raw,
uncorrected data (same data as in the selection boxes in
Fig.~\ref{fig:fullCMD} and sum of upper and lower panels of
Fig.~\ref{fig:radCMD}).  The middle panels show the CMDs after correction for
differential extinction/reddening using $E(B-V)$ values from the Schlegel
et~al.\ (1998) dust map, scaled to $A_V$ and $E(V-I)$ using the best-fit
$R_V=3.0$ and the corresponding $R_I$.  The mean extinction and reddening are
added back to the data to allow direct comparison to the uncorrected data.
The data in the bottom panels are corrected for differential visual
extinction on a star-by-star basis using the mean of the $\delta{A_V}$ values
from the ($B-V$,~$V$) and ($V-I$,~$V$) CMDs (\S\,4.1.1); this is scaled to a
color excess using the best-fit $R_V$ and $R_I$ values.  The tightness of the
resulting HB demonstrates the power of the ``$A_V$~RMS'' and ``$A_V$~Slope''
methods and the smoothing out of small-scale reddening variations in the SFD
dust map (see \S\,4.4).
}
\end{figure}

There is a handful of stars at each extremity of the HB in the bottom
two~panels of Figure~\ref{fig:cmd6} that do not follow the tight sequence
delineated by the rest of the stars---in fact, their scatter is unchanged
relative to the uncorrected CMDs in the top panels.  This is because the
polynomial HB fiducial used to determine $\delta{A_V}$ does not extend all
the way to the ends of the sequence (to avoid extrapolation of the fit); as a
result no $\delta{A_V}$ measurement or correction is available for these
stars.  A couple of stars near the middle of the HB continue to show large deviations in the 
$\delta{A_V}$-corrected CMDs but these are probably non-HB stars (see
\S\,4.1.1 and Fig.~\ref{fig:radCMD}).

Before closing out this discussion, two additional sources of HB scatter are
considered.  The first of these is photometric error.  Errors in $B$ and $I$
magnitudes produce uncorrelated shifts between the two~CMDs.  Errors in $V$
produce shifts with slope $-1$ in the ($B-V$,~$V$) CMD, more or less
orthogonal to the reddening vector, and shifts of slope $+1$ in the
($V-I$,~$V$) CMD, roughly along the reddening vector though its slope $R_I$
is significantly steeper than unity.  Only at the very top of the HB where it
becomes horizontal can $V$ magnitude errors mimic differential
extinction/reddening effects.  To summarize, photometric errors cannot
explain the correlated scatter seen in NGC~4833's HB.  Moreover, as will be
demonstrated in \S\,5, the difference in $\delta{A_V}$ between pairs of HB
stars scales with their angular separation in a manner consistent with the
dust power spectrum measured at larger angular separations; photometric
errors have no natural way of explaining this trend either.

Another factor to consider is {\it intrinsic\/} scatter in the HB caused by
astrophysical effects such as variations in chemical abundance, age, and mass
loss between cluster stars.  It is possible that these can produce correlated
scatter between the two~CMDs and thereby mimic differential
extinction/reddening.  However, this scenario cannot explain the scaling of
the $\delta{A_V}$ differences with angular separation (\S\,5).  If there was
a radial gradient in intrinsic stellar properties within the cluster, one
might expect to see a shift of the HB sequence in the CMD between inner and
outer regions.  Instead, Figure~\ref{fig:radCMD} shows no such shift (the
increased scatter in the HB in the outer CMD is best explained in terms of
dust variations; see \S\,2.1).

\section{Angular Power Spectrum of the Cirrus}     

\begin{figure}
\includegraphics[scale=0.7]{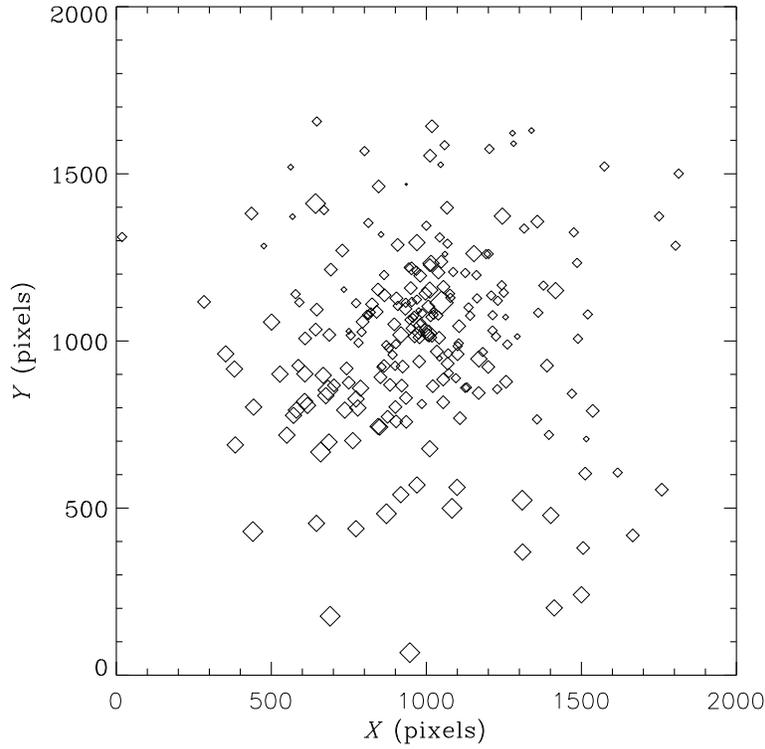}
\figcaption[cirrus3.ps]{\label{fig:dustmap}
A map of the {\it relative\/} visual extinction in a $13.5'\times13.5'$
region of the sky centered on NGC~4833.  Each diamond represents the sky
position of a cluster HB star.  The symbol size scales with
$(\delta{A_V})^{BV}$ [$-0.20$ (smallest) to $+0.37$ (largest)], the vertical
distance in the CMD between the HB star and the point where the reddening
vector drawn from the star intersects the fiducial (\S\,4.1.1).  The pixel
scale for $X$ and $Y$ axes is $0.396\arcsec\,$pix$^{-1}$.  North is along
$+Y$ and east is along $+X$.
}
\end{figure}

In earlier sections, we demonstrated the effects of spatial
extinction/reddening variations on the photometry of NGC~4833.  We now turn
to a statistical description of the strength of these extinction/reddening
variations as a function of angular scale.  Figure~\ref{fig:dustmap} is a 
map of the {\it relative\/} visual extinction across the face of NGC~4833.
Each point on the map represents the sky position of a HB star in the
cluster.  The size of each point is scaled by the measured strength of the
relative visual extinction $(\delta{A_V})^{BV}$ at that location; it varies
from $-0.20$ to $+0.37$~mag (smallest$\,\rightarrow\,$largest).  As described
in \S\,4.1.1, $(\delta{A_V})^{BV}$ is the vertical distance between the
star's CMD location and the point where the reddening vector drawn from the
star intersects the fiducial HB; the fiducial is a fit to the central region
of the cluster, and therefore corresponds to a mean visual extinction of
$\langle{A_V}\rangle\sim1$.  In keeping with the findings of Melbourne
et~al.\ (2000), there is a clear overall N-S gradient in $(\delta{A_V})^{BV}$
across NGC~4833, but the extinction can vary significantly from point to
point even for points relatively close to each other.

These variations can be characterized in terms of the angular power spectrum
of the foreground dust complex.  A direct measurement of the power spectrum
is difficult however because of the sparse and irregular spatial sampling of
the data points.  The following approach is adopted instead.  First, the
``variance spectrum'' as a function of angular scale is calculated from the
cluster $(\delta{A_V})^{BV}$ data (\S\,5.1).  Next, a variance spectrum and
conventional power spectrum are measured from the uniformly-sampled SFD maps,
and an empirical transformation is calculated from the former to the latter
(\S\,5.2).  Finally, this empirical transformation is applied to the cluster
variance spectrum to convert it to a power spectrum, and this allows us to
study the composite power spectrum over a wide range of angular scales
(\S\,5.3).

\subsection{Variance versus Angular Separation from NGC~4833 Optical Data}

\begin{figure}
\includegraphics[scale=0.7]{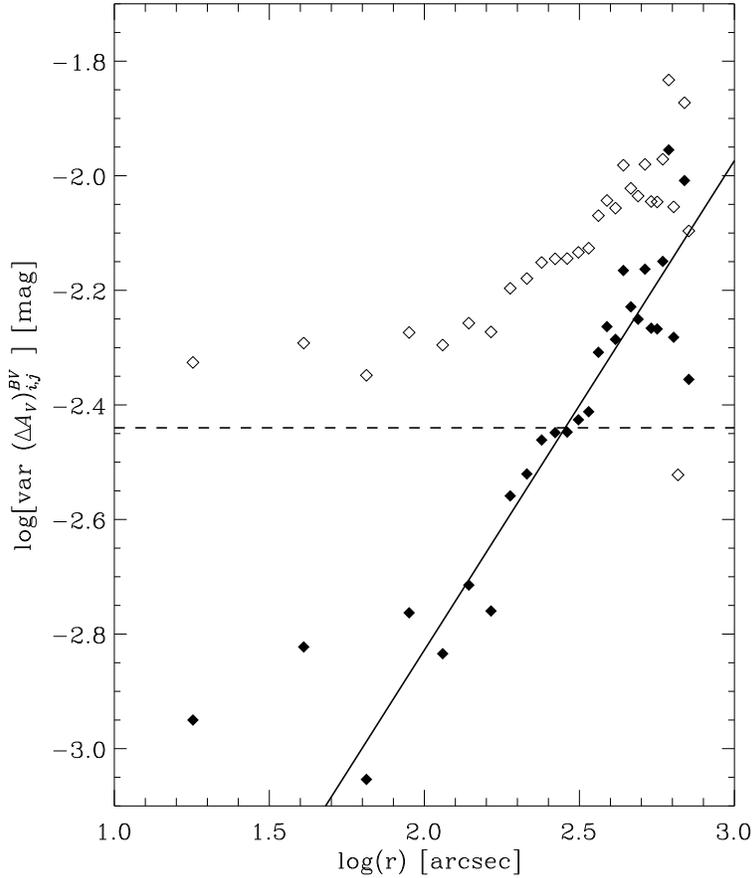}
\figcaption[cirrus3.ps]{\label{fig:powspec}
The variance of $(\Delta{A_V})^{BV}_{ij}$, the difference in relative visual
extinction between pairs of NGC~4833 HB stars $i$ and $j$ measured off the
($B-V$,~$V$) CMD (\S\,4.1.1), as a function of their projected separation
$r_{ij}$ (open diamonds).  The dashed horizontal line shows the expected
amount of variance from photometric errors alone, as derived from the special
``$\sigma_{\rm phot}$-only'' simulations that match the mean
extinction/reddedning and photometric errors in the cluster data set but
contain {\it no\/} reddening variations (\S\,3.3.1).  The filled diamonds
show the ``variance spectrum'' of NGC~4833 after removing the effects of
photometric error (\S\,5.1).  The solid line power-law fit to the corrected
variance spectrum has an index $\beta=+0.9\pm0.1$.
}
\end{figure}

The best-fit value of $R_V=3.0$ for NGC~4833 is used to determine the
relative visual extinction corresponding to the $i$-th HB star,
$(\delta{A_V})^{BV}_i$.  The difference in relative $A_V$ between stars $i$
and $j$,
$(\Delta{A_V})^{BV}_{ij}\equiv(\delta{A_V})^{BV}_i-(\delta{A_V})^{BV}_j$, 
and their angular separation $r_{ij}$ are then calculated for all possible
pairs of HB stars.  Figure~\ref{fig:powspec} (open diamonds) shows the
variance (mean square deviation) of $(\Delta{A_V})^{BV}_{ij}$ as a function
of angular separation $r$ in a log-log plot; the radial bins are chosen so as
to include roughly equal numbers of pairs in each bin.  The variance
decreases with decreasing $r$ before levelling off at log$\Bigl[{\rm
var}[(\Delta{A_V})^{BV}_{ij}]\Bigr]=-2.3$ for $r\lesssim100''$.  The behavior
at low $r$ may be attributed to photometric errors and possibly the natural
width of the HB.  

In order to quantify and compensate for the effect of photometric error on
the variance spectrum, we resort to the special set of ``$\sigma_{\rm
phot}$-only'' simulations containing {\it no\/} reddening variations
(\S\,3.3.1).  The variance of this ``$\sigma_{\rm phot}$-only'' simulated
data set, analyzed in the same way as the real cluster data set, is:
log$\Bigl[{\rm var}[(\Delta{A_V})^{BV}_{ij}]\Bigr]=-2.4$, and is of course
independent of stellar angular separation (dashed horizontal line in
Fig.~\ref{fig:powspec}).  The variance due to photometric errors is
subtracted from the variance spectrum of NGC~4833, equivalent to statistical
subtraction of errors in quadrature.  The resulting variance spectrum is
shown as filled diamonds in Figure~\ref{fig:powspec}.  A power-law fit to the
corrected variance spectrum:
\begin{equation}
{\rm var}(r)\propto{r}^{\beta},
\end{equation} 
yields an index of $\beta=+0.9\pm0.1$, over the radial range $50''$--$300''$.
The upper end of this range corresponds to the Nyquist frequency of the
$13.5'\times13.5'$ area covered by the NGC~4833 data set.  The first
radial bin of the corrected variance spectrum ($r<50''$) is excluded from the
power-law fit because it is strongly affected by uncertainties in the
photometric error correction.

The cluster variance spectrum is used to make a rough estimate of how much
small-scale power is missing from the SFD dust map.  The SFD map, like the
IRAS 100$\,\mu$m data on which they are based, have an angular resolution of
about $6'$, or ${\rm log}(r)\sim2.6$.  This point in the cluster variance
spectrum corresponds to var$[(\Delta{A_V})^{BV}_{ij}]=4.4\times10^{-3}$.  The
RMS scatter at this point is $\sqrt{{\rm
var}[(\Delta{A_V})^{BV}_{ij}]}=6.6\times10^{-2}$.  Since the mean visual
extinction $\langle{A_V}\rangle\approx1$ in the direction of NGC~4833, the
RMS is about 7\% of the mean.  This provides a good sanity check: it
corroborates the result from the analysis of the cluster HB width (\S\,3.3.2)
that the SFD map is missing small-scale variations on the order of 6\% of the
reported reddening value.  It also gives an indication of how accurately one
can correct for extinction/reddening using the SFD map, with general
implications for precision photometry in dusty regions of the sky.

\subsection{Variance and Power Spectra from SFD Map}

\begin{figure}
\includegraphics[scale=0.7]{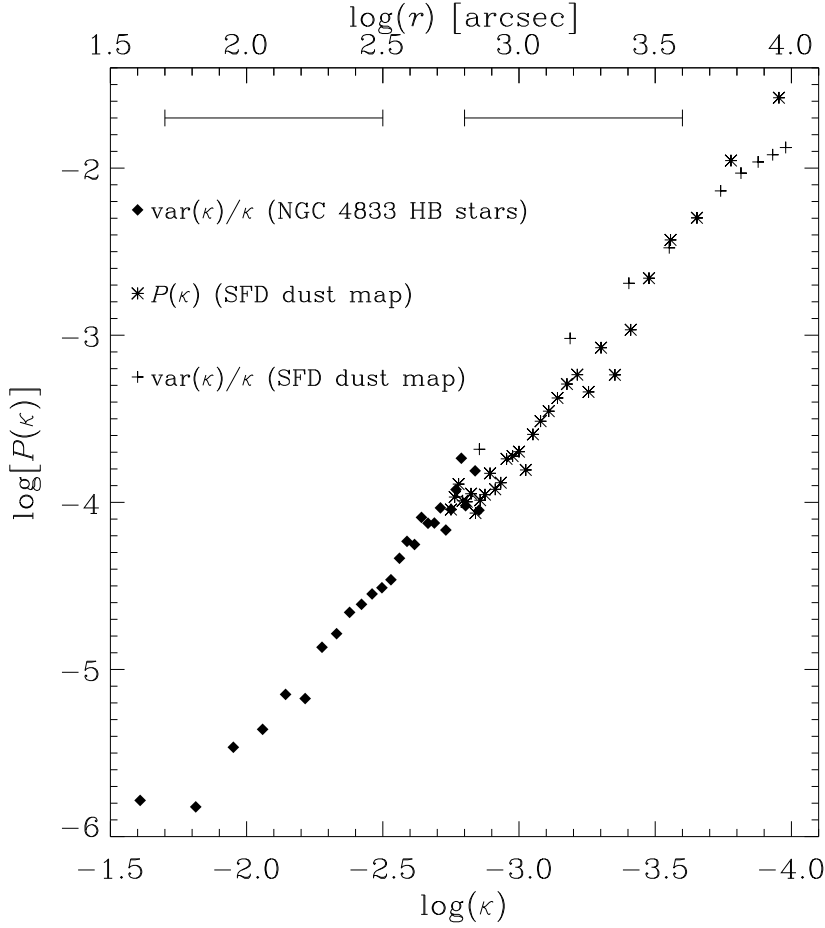}
\figcaption[cirrus3.ps]{\label{fig:pwvr}
The one-dimensional angular power spectrum $P(\kappa)$ of dust in the
foreground of NGC~4833, where $\kappa$ is the reciprocal of the projected
separation $r$.  The asterisks are measured from a $10^\circ\times10^\circ$
Schlegel et~al.\ (1998) reddening map centered on the cluster (\S\,5.2).  The
`$+$' symbols show ${\rm var}(\kappa)/\kappa$ from the SFD map, where the
variance of the difference in $E(B-V)^{\rm SFD}$ between pairs of pixels is
calculated for all pixel pairs as a function of their angular separation.
Note the general agreement between the SFD $P(\kappa)$ and ${\rm
var}(\kappa)/\kappa$.  The filled diamonds show ${\rm var}(\kappa)/\kappa$
(equivalent to the angular power spectrum) derived from the ($B-V$,~$V$) CMD
of NGC~4833 HB stars, corrected for the effects of photometric error (same
data as the filled diamonds in Fig.~\ref{fig:powspec} but with a different
binning in $r$).  The best-fit power-law indices are: $\alpha=-1.9\pm0.1$
over the range $r\sim1'$--$5'$ for optical cluster HB data, and
$\alpha=-2.0\pm0.1$ over the range $r\sim10'$--$1^\circ$ for the SFD map
(radial ranges indicated by horizontal bars near the top).
}
\end{figure}

In this section, the SFD map is used to investigate the relationship between
the variance spectrum and conventional angular power spectrum.  This is done
by extracting a $10^\circ\times10^\circ$ SFD reddening image centered on
NGC~4833.  The $E(B-V)^{\rm SFD}$ reddening values are converted to $A_V$
using the best-fit $R_V$ of 3.0.  The difference in visual extinction between
pixels $i$ and $j$, $(\Delta{A_V})^{\rm SFD}_{ij}\equiv(\Delta{A_V})^{\rm
SFD}_i-(\Delta{A_V})^{\rm SFD}_j$, is calculated for all possible pairs of
pixels.  The variance of $(\Delta{A_V})^{\rm SFD}_{ij}$ is computed as a
function of angular separation $r$.  The `$+$' symbols in
Figure~\ref{fig:pwvr} show ${\rm var}(\kappa)/\kappa$ for the SFD map, where
$\kappa\equiv1/r$.

The two-dimensional power spectrum of dust fluctuations is computed via a
Fast Fourier Transform (FFT) of the SFD reddening image.  This is then
azimuthally averaged to produce a one-dimensional angular power spectrum
$P(\kappa)$, shown by asterisks in Figure~\ref{fig:pwvr}.  Over the radial
range $r=10'$ to $1^\circ$ [${\rm log}(\kappa)\sim-2.8$ to $-3.6$], ${\rm
var}(\kappa)/\kappa$ and $P(\kappa)$ have comparable logarithmic slopes
(power-law index: $\alpha\approx-2.0\pm0.1$) such that:
\begin{equation}
\label{eqn:var_to_ps}
{\rm var}(\kappa)/{\kappa} \approx C\,P(\kappa)
\end{equation}
with a best-fit scale factor of ${\rm log}(C)=-4.3$.

\subsection{Composite Power Spectrum of the Cirrus}

The above transformation (Eqn.~\ref{eqn:var_to_ps}) is applied to the
photometric-error-corrected variance spectrum derived from the NGC~4833 $BV$
data (\S\,5.1) to convert it to an angular power spectrum.  The result is
shown as filled diamonds in Figure~\ref{fig:pwvr}.  This small-scale power
spectrum derived from optical data matches smoothly onto the power spectrum
derived from the SFD map at larger angular scales.  A power law with an index
$\alpha\sim-2.0$ provides an adequate fit to the data over the full range of
projected separations shown in Figure~\ref{fig:pwvr}, $r\sim1'$ to $3^\circ$.

Our best-fit index $\alpha$ is significantly shallower than the index of $-3$
found by Gautier et~al.\ (1992) and Guhathakurta \& Cutri (1994) and at the
shallow end of the range of indices found by Kiss et~al.\ (2003) in their
ISOPHOT-based study of 13~fields: $\alpha\sim-2.1$ to $-5.3$ ($\pm0.1$).  The
shallower the power-spectrum index, the larger the {\it fraction\/} of power
missed by the SFD map on small scales relative to that on larger scales.
Another possible implication relates to Kiss et~al.'s finding that shallower
$\alpha$ values are associated either with low HI column density, $N_{\rm
HI}<10^{21}$~cm$^{-2}$, and/or warm dust.  For example, in the region of the
Draco Nebula, they find that the map of 90$\,\mu$m emission (dominated by
warmer dust) yields $\alpha=-2.5\pm0.2$ while the 170$\,\mu$m map (dominated
by cooler dust) yields $\alpha=-4.1\pm0.2$.

\section{Future Work}

While NGC~4833 has an extended blue HB, the methods presented in this paper
should be usable for clump HBs or the RGB or any other tight CMD feature for
that matter.  For instance, Law et~al.\ (2003) used the width of the RGB to
measure differential reddening in the direction of three~Galactic globular
clusters, von Braun et~al.\ (2002) used the main sequence for their study,
and Udalski (2003) used clump stars to probe dust along the line of sight to
the Galactic bulge.  Our methods should work for any three~bands, not just
$BVI$.  Additional bands (beyond three) can provide: (1)~a longer wavelength
baseline and therefore more leverage in $R_V$ determination; (2)~a larger
number of independent measures of the star-by-star extinction/reddening which
can be averaged to beat down the errors; and (3)~most importantly, an
empirical check of the CCM parameterization of the shapes of Galactic
extinction laws.

There is at least one~generalization of the methodology presented here that
might be worth exploring in the future.  In determining $\delta{A_V}$, we
effectively use $V$ magnitude to mark where along the HB sequence the
extinction/reddening-corrected star lies.  A more general definition of a
``locator'' parameter, to mark the position of the unreddened star along any
given CMD sequence, may be particularly useful.  For example,
multi-wavelength data sets may not have any single observable in common
across all of the CMDs in question (unlike the $V$ mag for our two~CMDs), and
a color parameter may be more sensitive than $V$ mag when using a
not-so-vertical CMD feature (such as the upper portion of NGC~4833's HB).

Several high-quality, multi-band (three or more), ground- and space-based
photometric data sets have long been available for many star clusters, so the
methods presented here can be readily implemented to probe the nature of dust
along many lines of sight.  These include Milky Way clusters, as well as
others near enough to allow clean photometry in relatively uncrowded fields,
such as {\it Hubble Space Telescope\/} images of clusters in, and even
somewhat beyond, the Magellanic Clouds.  Many of these clusters have
detectable extinction/reddening variations across
them---e.g.,~$\omega$~Centauri, NGC~6388, NGC~6441 (Law et~al.\ 2003), M10,
and M12 (von Braun et~al.\ 2002).  The Sloan Digital Sky Survey (SDSS,
Stoughton et~al.\ 2002) five-band ($u'g'r'i'z'$) data set, with its high
photometric accuracy, minimal systematic errors, excellent
uniformity/homogeneity, and large areal coverage, is opening up lines of
sight towards differentially-reddened star clusters across the entire
high-latitude North Galactic cap for this type of analysis.  Near-infrared
$JHK$ photometry from the Two~Micron All-Sky Survey (2MASS) data should also
prove very fruitful, both for probing highly-reddened lines of sight and
for extending the SDSS wavelength coverage.  Making $R_V$ and power spectrum
measurements along many sight-lines should lead to a better understanding of
grain formation and chemistry and the effect of interstellar dust on
astronomical observations.

While our $3\times3$ grid of simulations covers some amount of parameter
space in photometric error and $R_V$, it is fairly limited in scope: it
cannot simply be translated to any arbitrary cluster data set.  Another data
set may be different from the NGC~4833 data set in terms of photometric
bandpasses, level of photometric error, areal coverage, detailed morphology
of CMD features, degree of extinction/reddening variations, and/or extinction
law slope.  New simulations must be carried out, matched to the exact
parameters of each new cluster data set in question, in order to test the
efficacy of the $R_V$ measurement methods presented here.

\section{Conclusions}

\begin{itemize}
\item[$\bullet$]{We have demonstrated the use of three~methods,
``$A_V$~RMS'', ``$A_V$~Slope'', and ``Optical/IR'' methods, for determining
the dust extinction law slope $R_V$ in the direction of
differentially-reddened Galactic globular clusters, and have tested the
methods on an extensive suite of simulated cluster data sets.}

\item[$\bullet$]{For cluster data sets with low photometric error,
$\sigma_{\rm phot}\lesssim0.01$~mag, the ``$A_V$~RMS'' and ``$A_V$~Slope''
methods can be used to determine $R_V$ to an accuracy of
$\sigma(R_V)=0.1$--0.3.  For cluster data sets with $\sigma_{\rm
phot}\gtrsim0.03$~mag, the two~methods yield $R_V$ to within $\pm0.7$, with
some systematic biases.}

\item[$\bullet$]{The ``Optical/IR'' method generally provides relatively
imprecise estimates of $R_V$, $\sigma(R_V)\sim0.6$, over the full range of
photometric errors explored.}

\item[$\bullet$]{Combining the results from all three~methods gives a mean
extinction law slope of $R_V=3.0\pm0.4$ for the line of sight towards the
low-latitute Galactic globular cluster NGC~4833.}

\item[$\bullet$]{The scatter in the cluster HB is used to estimate the amount
of small-scale structure in the dust complex in the foreground of NGC~4833.
The Schlegel et~al.\ (1998) IRAS+DIRBE-based map of the dust thermal emission
averages over small-scale reddening variations that are $\approx6$\% of the
mean reddening value.}

\item[$\bullet$]{Star-to-star variations in relative visual extinction across
the face of NGC~4833 provide a measure of the foreground dust angular power
spectrum for projected separations in the range $r\sim1'$--$5'$.  This
small-scale power spectrum derived from cluster optical data matches smoothly
onto the larger-scale power spectrum derived from the SFD reddening map of
the region.  The overall power spectrum is well fit by a power law:
$P(\kappa)\propto\kappa^\alpha$, where $\kappa$ is the reciprocal of the
angular scale $r$, with spectral index $\alpha\approx-2.0\pm0.1$.}
\end{itemize}

\acknowledgements
We would like to thank Neil Balmforth, David Burstein, Sandra Faber, and
Peter Stetson for comments that significantly improved this work.  We would
also like to thank Ata Sarajedini for providing the observations of NGC~4833.

\newpage

\newpage 
\vspace{-5cm}
\begin{deluxetable}{cc|c|ccc}
\tablecaption{\label{tbl:sim} $R_V$ Measurements}
\tablecolumns{6}
\tablehead{
Typical     & \% of       & True~$R_V$ & \multicolumn{3}{c} {Measured $R_V$
(Output) $\pm$ RMS Scatter}\\
Phot.~Error & NGC~4833    & (Input)    & $A_V$~RMS & $A_V$~Slope & Optical/IR \\
(mag)       & Phot.~Error &            & Method    & Method      & Method}
\startdata
\\[-0.9cm]
\cutinhead{Monte Carlo Simulations}
      &     & 2.0 &  $2.05\pm0.05$      &  $1.98\pm0.04$      & $2.10\pm0.68$\\ 
0.003 &  10 & 3.0 &  $3.09\pm0.09$      &  $2.97\pm0.09$      & $2.88\pm0.75$\\ 
      &     & 4.0 &  $4.06\pm0.12$      & $~4.29\pm0.18$\tablenotemark{a} & $3.97\pm0.82$\\
\hline
      &     & 2.0 &  $2.00\pm0.26$      &  $2.04\pm0.20$      & $1.99\pm0.64$\\
0.015 &  50 & 3.0 &  $2.97\pm0.42$      &  $3.03\pm0.45$      & $2.88\pm0.69$\\
      &     & 4.0 & $~3.68\pm0.50$\tablenotemark{a} &  $4.13\pm0.51$      & $3.92\pm0.81$\\
\hline
      &     & 2.0 &  $2.08\pm0.68$      & $~2.19\pm0.36$\tablenotemark{a} & $1.97\pm0.57$\\
0.030 & 100 & 3.0 & $~2.60\pm0.70$\tablenotemark{a} &  $2.95\pm0.60$      & $2.89\pm0.73$\\
      &     & 4.0 & $~3.13\pm0.92$\tablenotemark{a} & $~3.69\pm0.79$\tablenotemark{a} & $3.87\pm0.91$\\
\hline
\\[-0.6cm]
\cutinhead{NGC~4833}
0.030 & --- & --- & $~2.5(\rightarrow2.9)\pm0.7$\tablenotemark{b}   & $2.6\pm0.6$   & $3.5\pm0.7$    
\enddata
\tablenotetext{a}{These cases display a bias in the mean measured (output)
$R_V$ (more than $\pm0.1$) that is significant relative to the error in the
mean [$\sigma(R_V)/\sqrt{100}$].}
\tablenotetext{b}{The measured $R_V$ for NGC~4833 using the ``$A_V$~RMS''
method has been corrected upwards by $+0.4$ to compensate for measurement
bias.}
\end{deluxetable}

\end{document}